\documentstyle[aps,preprint,axodraw,psfig]{revtex}
\tighten
\newcommand{\gsim}{ \mathop{}_{\textstyle \sim}^{\textstyle >} }
\newcommand{\epem}{e^+e^-}

\newcommand{\tautomug}{\tau\to \mu\gamma}
\newcommand{\tautoeg}{\tau\to e\gamma}
\newcommand{\mutoeg}{\mu\to e\gamma}
\newcommand{\taumugamma}{\tau\to \mu\gamma}
\newcommand{\tauegamma}{\tau\to e\gamma}
\newcommand{\muegamma}{\mu\to e\gamma}
\newcommand{\tautoenunu}{\tau^-  \to e^-   \nu_\tau \bar{\nu}_e}
\newcommand{\tautomununu}{\tau^- \to \mu^- \nu_\tau \bar{\nu}_\mu}
\newcommand{\sla}[1]{\not\!#1}
\newcommand{\E}{ \sla{E} }
\newcommand{\LSP}{\tilde{\chi}^0_1}
\newcommand{\neutralinoone}{\tilde{\chi}^0_1}
\newcommand{\neutralinotwo}{\tilde{\chi}^0_2}
\newcommand{\charginoone}{\tilde{\chi}^-_1}
\newcommand{\charginoonep}{\tilde{\chi}^+_1}
\newcommand{\charginoonem}{\tilde{\chi}^-_1}
\newcommand{\charginoonepm}{\tilde{\chi}^\pm_1}

\newcommand{\seL}{\tilde{e}_L}
\newcommand{\smuL}{\tilde{\mu}_L}
\newcommand{\stauL}{\tilde{\tau}_L}

\newcommand{\stauR}{\tilde{\tau}_R}
\newcommand{\stau}{\tilde{\tau}}
\newcommand{\slepL}{\tilde{l}_L}

\newcommand{\snu}{\tilde{\nu}}
\newcommand{\sneu}{\tilde{\nu}}
\newcommand{\snue}{\tilde{\nu}_e}
\newcommand{\snumu}{\tilde{\nu}_\mu}
\newcommand{\snutau}{\tilde{\nu}_\tau}

\newcommand{\sneutau}{\tilde{\nu}_\tau}
\newcommand{\e}{e}
\newcommand{\Mgrav}{M_{\rm grav}}

\begin{document}
\draft
\preprint{
\begin{minipage}[t]{3in}
\begin{flushright}
KEK-TH-675, UT-867 \\
hep-ph/0004256 
\end{flushright}
\end{minipage}
}
\title{Probing Left-handed Slepton Flavor Mixing\\
 at Future Lepton Colliders}
\author{Daisuke Nomura
  \thanks{JSPS Research Fellow, 
          e-mail address:{\tt  daisuke.nomura@kek.jp}}}
\address{
Department of Physics, University of Tokyo, Tokyo 113-0033, Japan\\
    and \\
Theory Group, KEK, Tsukuba, Ibaraki 305-0801, Japan} 
\date{revised in June, 2001}
\maketitle
\begin{abstract}
\setlength{\baselineskip}{17pt}
It has been argued in the literature that the search for
the slepton oscillation phenomenon can be a powerful probe 
of intergenerational mixing between sleptons, 
once sleptons are found at future colliders.
In this article we estimate possible reach of future lepton colliders
in probing left-handed slepton flavor mixing, 
especially mixing between the first and third generations,
on which constraints imposed by other processes like  
$\tau \to e \gamma$ are very weak.
$e^+e^-$ collider is suitable for this purpose, since
it can produce, if kinematically allowed, 
sleptons of the first generation via $t$-channel, 
in addition to $s$-channel. 
Utilizing $e^+e^- \to \tau e + 4jets + \E$ signal 
at $e^+e^-$ linear collider with integrated luminosity 
${\cal L}=50 {\rm fb}^{-1}(500 {\rm fb}^{-1})$ it may be possible to reach 
mixing angle $\sin 2\theta_{\tilde{\nu}} \gtrsim 0.06 (0.04)$ and 
mass difference $\Delta m_{\tilde{\nu}}  \gtrsim 0.07 (0.04)$ GeV 
for sneutrinos in the first and third generations 
at the statistical significance of 5$\sigma$.
\end{abstract}
\pacs{PACS numbers: 12.60.Jv, 11.30.Hv, 14.80.Ly}

\section{Introduction}
%\label{sec:intro}

Introduction of the weak scale supersymmetry (SUSY) 
into the Standard Model (SM) provides 
an elegant solution to the naturalness problem.
In this scenario SUSY dictates cancellation of quadratic divergences
between the fermionic and bosonic loop contributions to
the radiative corrections to the Higgs boson mass-squared,
which naturally keeps the Higgs boson as light as the weak scale
against the radiative corrections.
However, it simultaneously causes other problems, 
including the flavor problem as one of the most serious ones. 
That is, while intergenerational mixing among sfermions is not determined 
unless one puts additional assumptions about their origin, 
it is often the case that 
arbitrary flavor-violating soft SUSY breaking parameters
result in unacceptably large rates in the 
Flavor Changing Neutral Current (FCNC) processes. 
In other words, the mechanism with which
the Nature avoids such a disaster 
may be directly linked to the origin of SUSY breaking,
and it is certain that experimentally sensitive probes 
will provide indispensable clues for it.

Another motivation for probing flavor mixing of sfermions, 
especially left-handed sleptons,
comes from recent results of neutrino oscillation 
experiments~\cite{SKatm,SKsolar}.
They suggest that the neutrinos have not only tiny masses
but also large angle mixing(s) among them.
This means that the flavor structure of 
the lepton sector may be totally different 
from that of the quark sector 
whose intergenerational mixings are relatively small.
In some of supergravity(SUGRA)-motivated SUSY models 
with the seesaw mechanism~\cite{seesaw},
large neutrino mixing 
suggested by the neutrino oscillation experiments
may be due to large flavor mixings 
in neutrino Yukawa coupling constant matrix 
which is responsible for the Dirac masses of neutrinos. 
In that case, via radiative corrections
they cause sizable lepton-flavor violating (LFV) elements 
in the mass-squared matrix of the left-handed slepton, 
which includes the superpartner of neutrino, 
sneutrino~\cite{LFVneutrino}.  
Hence probing the intergenerational mixing among 
sneutrinos may provide some hints for the origin of 
the flavor structure of the lepton sector 
or that of the SUSY breaking.
In this article we investigate 
the slepton oscillation phenomenon~\cite{Krasnikov,ACFH1}
between left-handed sleptons as one of such possible probes
with $e^+ e^-$ linear colliders (LC's).

The slepton oscillation is a quantum-mechanical phenomenon, 
analogous to the neutrino oscillation, induced by discrepancy
between gauge eigenstates and mass eigenstates.
If there are LFV elements in the slepton mass-squared matrix,
gauge eigenstates do not coincide with mass eigenstates.
Then a selectron produced at a collider, for example, 
will change its flavor (``oscillate'') into a stau
with some probability 
and decay into a tau lepton possibly with other superparticle(s).
Since the probability 
with which the slepton changes its flavor into other
is determined by the flavor mixing of sleptons,
from the relevant cross section of the LFV signal
we can obtain information  
to what extent sleptons of different flavors 
mix each other.

The slepton oscillation as a means of detecting slepton flavor mixing
\footnote{
Another interesting application of the slepton oscillation
is as a probe of a possible CP violating phase 
in the slepton sector~\cite{ACFH2}.
In this article we restrict ourselves in the case where
the slepton mass matrix is real, for simplicity.} 
has a unique advantage over other loop-induced 
(charged-)lepton-flavor violating 
processes~\cite{LFVneutrino,LFVGUT}, 
such as $\mutoeg$, $\mu\to 3e$, $\mu$-$e$ conversion in nucleus, 
$\tautomug$, $\tautoeg$, and so on.
That is, since real sleptons are produced in the former processes, 
GIM-like suppression is weaker than the latter ones.
For instance, in the case where there is a mixing only between 
the first and second generations, 
the cross section $\sigma$ of the typical signal 
of the slepton oscillation that contains $e$ and $\mu$ 
in the final state
depends on the masses $m_1$ and $m_2$ of the selectron $\tilde{e}$ 
and the smuon $\tilde{\mu}$, 
the mixing angle $\theta$ between them, and
the decay width $\Gamma$ of them as
\begin{eqnarray}
\sigma & \propto & \sin^2 2\theta 
           \frac{(\Delta m^2)^2}{4 \bar{m}^2 \Gamma^2 + (\Delta m^2)^2}, 
\end{eqnarray}
where $\Delta m^2 = m_1^2 - m_2^2$, and $\bar{m} = (m_1+m_2)/2$,
(Here we have taken the decay width $\Gamma$ as common between 
$\tilde{e}$ and $\tilde{\mu}$, for simplicity.), 
while a typical suppression factor in 
loop-induced processes, such as $\mutoeg$, is 
$\sin 2\theta \Delta m^2 / \bar{m}^2 $ for their amplitudes. 
In the minimal SUGRA scenario~\cite{Nilles}
with reasonable parameters typical decay width of slepton 
is ${\cal O}(1)$ GeV, 
it follows that the GIM suppression in the slepton oscillation is weaker
for small $\Delta m^2$.
Whether this directly means that the slepton oscillation is really
a sensitive probe or not is of course non-trivial 
and needs further investigation. 
We need to take into account 
competition against the back ground (BG) processes, 
effects of cuts to distinguish the signal out of the BG's, and so on.
Nevertheless, it has been reported in the 
literature~\cite{Krasnikov,ACFH1,HirouchiTanaka,HNST,AgasheGraesser,CHZL} 
that the slepton oscillation is indeed a powerful tool 
once the sleptons are found at future collider experiments.

The sensitivity of the slepton oscillation 
as a probe of slepton flavor mixing 
depends on the generations whose mixing is to be probed.
Comparing the sensitivity of the slepton oscillation studied in the 
literature~\cite{Krasnikov,ACFH1,HirouchiTanaka,HNST,AgasheGraesser,CHZL} 
with the current experimental bounds for 
charged-lepton flavor violating processes,
\begin{eqnarray}
{\rm Br}(\mutoeg)   &<& 1.2 \times 10^{-11}~\cite{MEGA}, \
                                            \label{muegamma} \\ 
{\rm Br}(\tautomug) &<& 1.1 \times 10^{-6}~\cite{CLEOtaumugamma},
                                            \label{taumugamma} \\
{\rm Br}(\tautoeg)  &<& 2.7 \times 10^{-6}~\cite{CLEOtauegamma},
                                            \label{tauegamma}
\end{eqnarray}
at least for their sample parameters 
the slepton oscillation is indeed a powerful tool 
that may go beyond these bounds.
However, comparing them with the reach of {\it future} experiments, 
it is not necessarily so.
For the mixing between the first and second generations
a proposed experiment
at PSI~\cite{PSIprop} is so sensitive that it will probe 
$\mutoeg$ decay to $10^{-14}$ branching ratio. 
If it is carried out, the sensitivity of it will reach
far beyond that of the slepton oscillation of the left-handed sleptons
between the first and second generations,
as discussed later in this article. 
(It can probe at most the parameter region that 
corresponds to $\mutoeg$ branching ratio of order of $10^{-12}$
for our sample parameters.)
On the other hand, in the case of 13-mixing and 23-mixing
\footnote{
From now on we refer to lepton flavor mixing 
between $i$-th and $j$-th generations 
as $ij$-mixing, for brevity.},
not only the current bounds (\ref{taumugamma}) and (\ref{tauegamma}) 
but also near future searches of these rare decay processes 
are not so powerful as in the 12-mixing case. 
While $\tautomug$ searches at the $B$ factories in KEK or SLAC 
may improve the bound to the level of $10^{-(7-8)}$,
according to the result of Ref.~\cite{HNST} 
the $3\sigma$ reach of LC in probing the left-handed slepton mixing
between the second and third generations
will go far beyond the reach of these near future $\tautomug$ searches
for their sample parameters.
Also, in the study of the slepton oscillation between 
the right-handed sleptons in Ref.~\cite{HirouchiTanaka} 
it is reported that
some of the LFV processes with tau lepton(s) in the final state
could have sizable cross sections at future linear collider experiments.
Hence the situation that the slepton oscillation can be 
the most valuable is in the case where we probe
slepton flavor mixing involving the third generation. 

In this article we discuss possible reach of future lepton colliders 
in probing flavor mixings of left-handed sleptons.
We probe them since they contain sneutrinos, 
the mixings of whose superpartners had not been revealed until recently.
We mainly focus on probing the 13-mixing of left-handed sleptons, 
which seems to be most unknown in neutrino mixing.
In probing it we also have an advantage that 
it may be more economically probed in an electron collider
rather than in a muon collider with the same energy and the same
luminosity, 
since in the former the $t$-channel pair production 
of the sleptons in the first generation,
which are relevant to the 13-mixing, 
is possible in addition to that via $s$-channel, if kinematically allowed.
Utilizing $e^+e^- \to \tau e + 4jets + \E$ signal 
at $e^+e^-$ linear collider with integrated luminosity 
${\cal L}=50 {\rm fb}^{-1}(500 {\rm fb}^{-1})$ 
we find it possible to reach 
mixing angle $\sin 2\theta_{\tilde{\nu}} \gtrsim 0.06 (0.04)$ and 
mass difference $\Delta m_{\tilde{\nu}}  \gtrsim 0.07 (0.04)$ GeV 
for sneutrinos in the first and third generations 
at the statistical significance of 5$\sigma$
for our sample parameter.

Contents of this article are as following.
In the next section we describe our framework within which
we discuss LFV at future colliders.
In section~\ref{sec:signalsandBGs} we discuss possible signals of LFV and
back ground processes for them, and estimate reach of future colliders
in probing the left-handed slepton flavor mixing between the first and 
third generations.
In section~\ref{sec:12-mixing} we discuss an application
to the left-handed slepton flavor mixing between the first and 
second generations, which is experimentally severely constrained by $\mutoeg$.
In section~\ref{sec:radiativemixing} we further apply our results in
section~\ref{sec:signalsandBGs} to the case in which 
the left-handed slepton flavor mixings are radiatively induced 
by right-handed neutrino Yukawa couplings.
Section~\ref{sec:conclusions} is devoted to 
our conclusions.

\section{Framework}
\label{sec:framework}

In this section we explicitly describe our framework
within which we analyze reach of the future lepton colliders
in probing the left-handed slepton flavor mixing.
The slepton oscillation between the left-handed sleptons 
is formerly considered in Ref.~\cite{HNST} in the case of 23-mixing.
Here we extend their result to other generations.
First, to make our point clearer and simpler, 
we confine ourselves in the case where
there is left-handed slepton flavor mixing 
only between the first and third generations.
Cases with other generation mixing will be briefly discussed later.
We define a mixing angle $\theta_{\snu}$ 
and a mass difference $\Delta m_{\snu}$ at the weak scale
that parameterize the left-handed slepton flavor mixing 
between the first and third generations
in the sneutrino mass-squared matrix $(m^2_{\snu})$, as
\begin{eqnarray}
(m^2_{\snu}) = 
\left(               
\begin{array}{rr}
\cos\theta_{\snu} & -\sin\theta_{\snu} \\ 
\sin\theta_{\snu} &  \cos\theta_{\snu} 
\end{array}
\right)
\left(               
\begin{array}{cc}
m^2_{\snue} & 0 \\ 0 & m^2_{\snutau}
\end{array}
\right)
\left(               
\begin{array}{rr}
 \cos\theta_{\snu} & \sin\theta_{\snu} \\ 
-\sin\theta_{\snu} & \cos\theta_{\snu} 
\end{array}
\right)
\end{eqnarray}
with $\Delta m_{\snu} \equiv m_{\snue} - m_{\snutau} $,
where the second generation is omitted 
and we are in the basis 
where the charged lepton mass matrix is diagonal
and no LFV in fermion-sfermion-chargino vertex or
in fermion-sfermion-neutralino vertex.
(Although if $\theta_{\snu}$ is non-zero 
the mass eigenstates are of course not equal to the weak eigenstates,
we refer to sneutrino masses as electron sneutrino mass $m_{\snue}$ 
and tau sneutrino mass $m_{\snutau}$ 
even for non-zero $\theta_{\snu}$ case, 
analogously to the $\theta_{\snu} \to 0$ limit.)
While there are many theoretical models that predict various values
for $\Delta m_{\sneu}$ and $\theta_{\sneu}$ at the weak scale, 
in this article, although we will choose sample parameters,
we do not specify any origin for the flavor structure.
We do our analysis to some extent model-independently, that is,
we simply assume that there is an intergenerational mixing
between the first and third generations, 
and treat $\Delta m_{\sneu}$ and $\theta_{\sneu}$ 
as free and independent parameters.

Next we choose our sample SUSY parameters.
To study all the potential of the colliders actually requires
full parameter search for all the unknown parameters, 
which is beyond our scope in this article.
Instead we choose representative parameter sets.
We choose almost the same parameters as those used in Ref.~\cite{HNST}.
These mass spectra are typical ones
in the minimal SUGRA scenario, and in Ref.~\cite{HNST} they 
actually assume the minimal SUGRA in order to determine 
masses of sfermions at the weak scale.
In this article, however, we do not necessarily assume the minimal SUGRA 
and simply regard these parameters as given.    

The numbers we used in our numerical calculation 
are listed in Table~\ref{MSSMparametertable}.
We fix the electron sneutrino mass $m_{\snue}$ 
at the weak scale as 180 GeV.
Masses of other sfermions at the weak scale are determined
by assuming the universal scalar mass at the gravitational scale
($\sim 10^{18}$GeV) and solving renormalization group equations (RGE's) 
of the minimal SUSY standard model (MSSM),
although once we obtain the sfermion mass spectra 
we no longer persist on the minimal SUGRA, as stated above.
The Higgsino mass parameter $\mu$ is determined 
from the radiative breaking conditions of the electroweak symmetry.
We fix the lighter chargino mass $m_{\charginoone}$ as 100GeV, and
assume the GUT relation on gaugino masses, 
\begin{equation}
\frac53 \frac{g_Y^2}{M_1} = \frac{g_2^2}{M_2} = \frac{g_3^2}{M_3}, 
\end{equation}
to fix other gaugino masses, 
although we do not necessarily assume the GUT. 
(Here $g_Y$, $g_2$, and $g_3$ are the gauge coupling constants
of the SM gauge group $U(1)_Y, SU(2)_L$, and $SU(3)_c$, respectively. 
$M_1, M_2$, and $M_3$ are B-ino, W-ino, and gluino masses, respectively.)
With these parameters we are in the region 
where the lightest SUSY particle (LSP) is 
the lightest neutralino $\LSP$, which is B-ino like, 
and the second-lightest neutralino $\neutralinotwo$ 
and the lighter chargino $\charginoonepm$ both W-ino like. 
In Ref.~\cite{HNST}, decay branching ratios of 
$\neutralinotwo$,  $\charginoonepm$ and the left-handed sleptons 
are calculated, and we will later use these values also in this article. 
(These values are listed in Table II and III.)

As for the colliders, we implicitly assume the use of the proposed LC,
like the JLC~\cite{JLC-I} or the NLC~\cite{NLC}, 
operated with the center-of-mass energy $\sqrt{s}=500$GeV.

For the statistical significance
we require that the significance ${\cal S}$ of the signal, defined as 
\begin{equation}
{\cal S} \equiv \frac{\sigma_{sig} {\cal L}}{\sqrt{\sigma_{bg} {\cal L}}},
\end{equation}
to be larger than 5 in order for the relevant LFV process
to be discovered at $5\sigma$ significance,
where $\sigma_{sig}$ and $\sigma_{bg}$ are 
the signal and background cross sections
after cuts (in fb), 
and ${\cal L}$ the integrated luminosity (in ${\rm fb}^{-1}$).

\section{signals and backgrounds for probing $\snu_\e$-$\snu_\tau$ mixing}
\label{sec:signalsandBGs}

In this section we discuss possible collider signals and BG's 
in probing the left-handed slepton flavor mixing between 
the first and third generations. 
Since decay channels of SUSY particles depend on
SUSY mass spectrum, to discuss all the possibility is 
too complex to be discussed in this article. 
We discuss them only within the framework 
described in the previous section.
\subsection{Signals}
First we have to decide what final state we should choose as signal.
The simplest choice would be 
$\epem \to \slepL^+ \slepL^- \to e^\pm \tau^\mp \LSP\LSP$, 
where $\slepL^\pm$ is the charged left-handed slepton 
with $l=e,\mu$ or $\tau$.
However since for our sample parameter set the left-handed sleptons are
heavy enough to decay into $\charginoonepm$ or $\neutralinotwo$, 
in addition to $\LSP$, and $\LSP$ is B-ino like 
while $\neutralinotwo$ and $\charginoonepm$ W-ino like,
the cross section of this mode is suppressed by 
the small decay branching ratio of the left-handed sleptons into $\LSP$,
as can be read off from Table~\ref{slepdecaytable}~\footnote
{If the decay $\slepL^\pm \to \neutralinotwo l^\pm$ 
is not kinematically allowed, 
left-handed sleptons can decay only as 
$\slepL^\pm \to \neutralinoone l^\pm$.
In that case the signal cross section 
$\sigma(\epem \to  \slepL^+ \slepL^- \to 
e^\pm \tau^\mp \LSP \LSP)$ is sizable.
However since this final state also results 
from pair production of right-handed sleptons 
if there are lepton-flavor mixings among right-handed sleptons, 
and telling the signal of left-handed slepton mixing from 
that of right-handed one is nontrivial and needs further investigation.
Here we confine ourselves in the case where
the decay $\slepL^\pm \to \neutralinotwo l^\pm$ is allowed 
and hence we can use the signal 
$\epem \to e^\pm \tau^\mp + 4jets + \E$, 
which we use as signal as described later, 
that can result from left-handed slepton pair production 
and cannot from right-handed one.}.
Assuming $t$-channel dominance, the cross section $\sigma$ is~\footnote
{Detailed formulas for evaluating amplitude and 
cross section on the slepton oscillation are found in 
Ref.~\cite{ACFH2}.}
\begin{eqnarray}
\sigma(\epem \to e^+ \tau^- \LSP \LSP) & \simeq &
 4 \chi_{\snu} \sin^2 2 \theta_{\snu} 
( 1 - \chi_{\snu} \sin^2 2\theta_{\snu} ) \times 0.09  ~{\rm fb},
\label{directLSPcrosssection} 
\end{eqnarray}
where $\chi_{\snu}$ is given as
\begin{eqnarray}
\chi_{\snu} &\equiv& 
\frac{ (x_{13}^{(\snu)})^2 }{ 2 ( 1 + (x_{13}^{(\snu)})^2 )}
\label{defofchi}
\end{eqnarray}
with
\begin{eqnarray}
x_{13}^{(\snu)} &\equiv& 
\frac{ m^2_{\snue} - m^2_{\snutau} }{ 2 [ m \Gamma ]_{13} } 
\end{eqnarray}
and
\begin{eqnarray}
[ m \Gamma ]_{13} & \equiv & 
( m_{\snue}   \Gamma_{\snue} + m_{\snutau} \Gamma_{\snutau} )/2 .
\end{eqnarray}
Here $\Gamma_{\snue}$ and $ \Gamma_{\snutau}$ are the decay widths
of electron sneutrino and tau sneutrino, respectively.
The SM BG of $e^\pm \mu^\mp$ + missing energy is estimated 
as 5.2 fb in the work of Arkani-Hamed et al.~\cite{ACFH1} 
for unpolarized beam, assuming 
the formerly studied efficient cuts~\cite{BeckerVanderVelde}
originally devised for flavor-conserving signal 
can be also effective to their case.
(Applying cuts of Ref.~\cite{BeckerVanderVelde},
$W$ pair production, $eW\nu$, and $(ee)WW$ contribute to the BG's 
1 or 2 fb each.)
If we also assume that in our  $e^\pm \tau^\mp$ case
we have the same order of BG as their $e^\pm \mu^\mp$ case,
the value of signal cross section of eq.~(\ref{directLSPcrosssection}) 
is too small to be observable. 
(In order for the signal of 0.1fb to be discovered 
at $5\sigma(3\sigma)$ significance in 1 year operation of LC,
the BG cross section must be smaller than 0.02(0.06)fb after applying cuts, 
even if we assume 100\% acceptance rate for signal.)
What makes matter worse, there is SUSY BG
from W-ino like chargino pair production in our case, 
in addition to SM BG's.
If we are to find out the right-handed slepton mixing
right-hand polarized beam can both reduce W-ino pair production rate
and enhance the signal rate at the same time,
as previously considered~\cite{ACFH1},
but now since we would like to probe the left-handed mixing, 
polarizing beam to right-handed will reduce both the signal and the 
BG simultaneously.
Therefore without any drastically better cuts this mode is not usable.
To find out such cut is beyond our aim, and we simply discard this mode.

Fortunately we can extract the information of LFV 
from the decay of sleptons into $\neutralinotwo$ or $\charginoonepm$.
That is,
\begin{eqnarray}
\epem &\to& \slepL^+ \slepL^-  ~~~{\rm or}~~~  \sneu\sneu^c   \nonumber  \\
      &\to& e^\pm \tau^\mp + 4jets + \E      \label{4jetsignal}
\end{eqnarray}
has sizable cross section. The 23-mixing version of this signal,
\begin{eqnarray}
\epem &\to& \slepL^+ \slepL^-  ~~~{\rm or}~~~  \sneu\sneu^c   \nonumber  \\
      &\to& \mu^\pm \tau^\mp + 4jets + \E,      \label{4jetsignalHNST}
\end{eqnarray}
is previously used in Ref.~\cite{HNST} in order to probe
the 23-mixing of the left-handed sleptons and turned to be efficient. 
In this mode each charged sleptons (sneutrinos)
decays into $\neutralinotwo$ ($\charginoonepm$) with a charged lepton, 
and each $\neutralinotwo$ ($\charginoonepm$) decays into 2 jets 
associated with $\LSP$, which we observe as missing energy.
Main advantage of this signal is that it is almost free from SM BG's,
since the final state contains 
4 jets and two charged leptons of different flavors~\cite{HNST}.
Another advantage is that in this mode sneutrino pair production
cross section contributes to the signal cross section, 
in addition to those from charged slepton.

\begin{figure}
\begin{center}
\leavevmode
\psfig{file=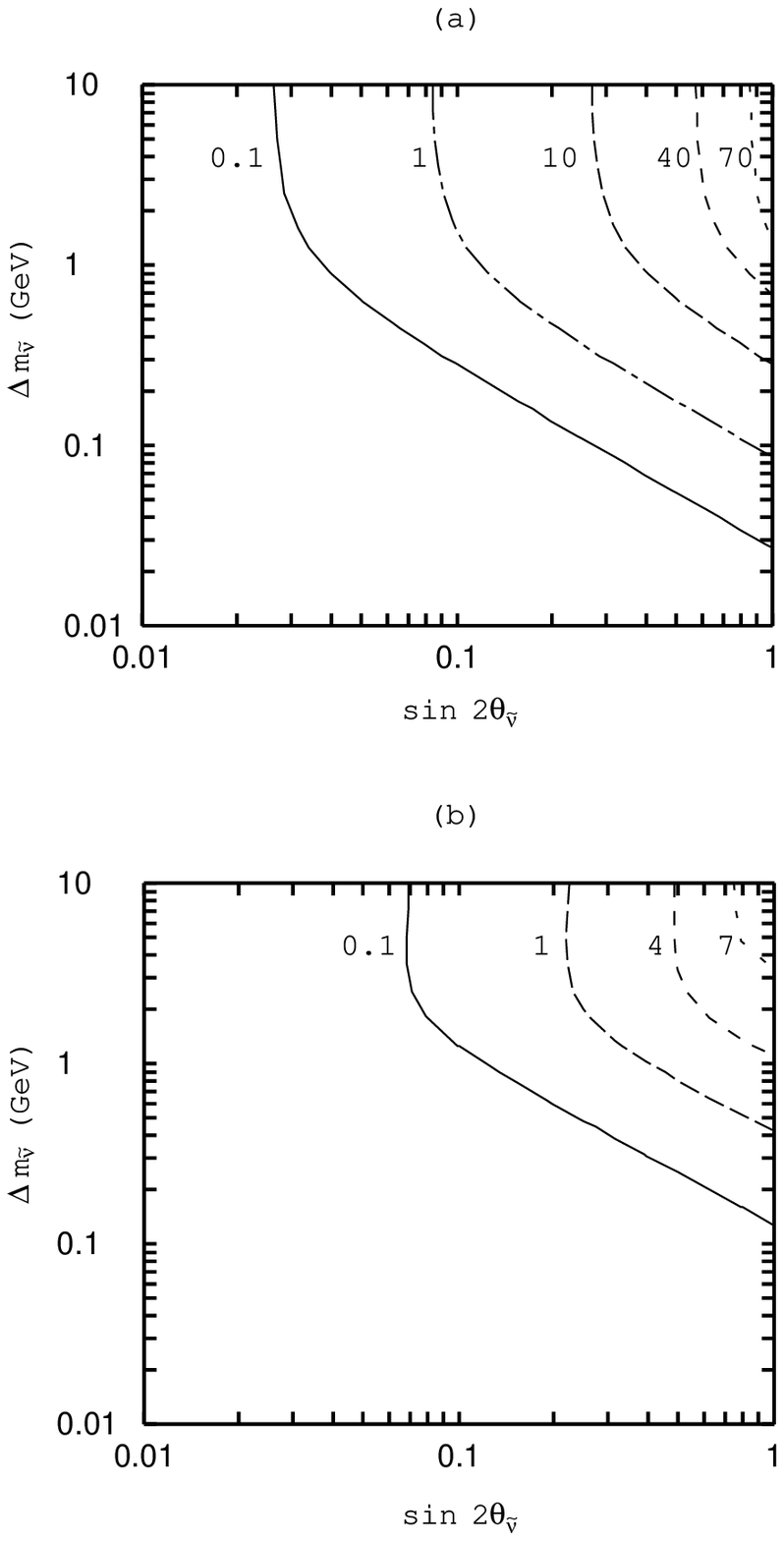,height=18cm}
\vspace*{1cm}
\caption{Contours of constant cross sections (in unit of fb) of 
 (a) $\epem \to \tau^+ e^- \charginoonep \charginoonem$ and 
 (b) $\epem \to \tau^+ e^- \neutralinotwo \neutralinotwo$ 
as functions of the mixing angle $\theta_{\snu}$ and 
the mass difference $\Delta m_{\snu}$
between the first and third generations
at the center-of-mass energy 500GeV.
We take $\tan\beta=3$ in the figures and other SUSY parameters 
are shown in TABLE I.}
\label{fig:sigmatanbeta03}
\end{center}
\end{figure}

We show contours of constant cross sections of 
$\epem \to \sneu\sneu^c \to \tau^+ e^- \charginoonep \charginoonem$
and 
$\epem \to \slepL^+ \slepL^- \to \tau^+ e^- \neutralinotwo\neutralinotwo$
in Figure~\ref{fig:sigmatanbeta03} (a) and (b), respectively,
as functions of the mixing angle $\theta_{\snu}$ and 
the mass difference $\Delta m_{\snu}$ between non-muon sneutrinos. 
In the figures we fix the electron sneutrino mass as 180GeV, 
the lighter chargino mass $m_{\charginoonepm}$ 100GeV, and $\tan\beta=3$,
and we determined other supersymmetric parameters by solving 
the RGE's of the MSSM, as described in the previous section.
Values of these parameters are summarized in Table I.
For these sample parameter
the cross section $\sigma$ of the process (\ref{4jetsignal}) is 
approximately
\begin{eqnarray}
\sigma(\epem \to e^+ \tau^- + 4jets + \E) & \simeq &
 4 \chi_{\snu} \sin^2 2 \theta_{\snu} 
( 1 - \chi_{\snu} \sin^2 2\theta_{\snu} ) \times 30  ~{\rm fb} 
\label{4jetcrosssection} 
\end{eqnarray}
with $\chi_{\snu}$ defined by eq.~(\ref{defofchi}).
(Here we have used decay branching ratios of SUSY particles 
calculated in Ref.~\cite{HNST}. 
This is reasonable since we had chosen our sample parameters
almost the same as theirs.)
We can have cross section of order of 10fb since 
in probing 13-mixing with $\epem$ collider we have an advantage 
that $t$-channel is also available, in addition to $s$-channel,
in order to produce left-handed selectrons. 
This is the point of this article.
In Ref.~\cite{HNST} since in their case 
only $s$-channel is available for $\epem$ collider to produce
the left-handed sleptons that are involved in generation mixing
(namely, left-handed sleptons in the second and third generation 
in their case), the cross section of (\ref{4jetsignalHNST}) is 
of order of 1fb.

From the equation~(\ref{4jetcrosssection}) 
we can understand the behavior of the contours as follows.
The cross section is related to the mixing angle $\theta_{\snu}$ 
and the mass difference $\Delta m_{\snu}$ approximately as
\begin{eqnarray}
\sigma & \propto & \sin^2 2\theta_{\snu}  
   \frac{(\Delta m_{\snu}^2)^2}{4 \bar{m}^2 \Gamma^2 + (\Delta m_{\snu}^2)^2}, 
\end{eqnarray}
where $\Delta m_{\snu}^2 \equiv m^2_{\snue}- m^2_{\snutau}$ 
and $\bar{m} = (m_{\snue} + m_{\snutau})/2$.
Therefore, for constant $\Delta m_{\snu}$ the cross section behaves as 
$\sigma \propto \sin^2 2\theta_{\snu}$. 
On the other hand, 
for a fixed value of $\theta_{\snu}$, $\sigma$ is proportional to 
$(\Delta m_{\snu}^2)^2$ 
in the region of $\frac{\Delta m_{\snu}^2}{\bar{m}} \ll \Gamma$, 
while $\sigma$ is almost independent of $(\Delta m_{\snu}^2)^2$
in the region that $\frac{\Delta m_{\snu}^2}{\bar{m}} \gg \Gamma$.
Hence we obtain characteristic curves 
for the contours of constant cross section.

\begin{figure}
\begin{center}
\leavevmode
\psfig{file=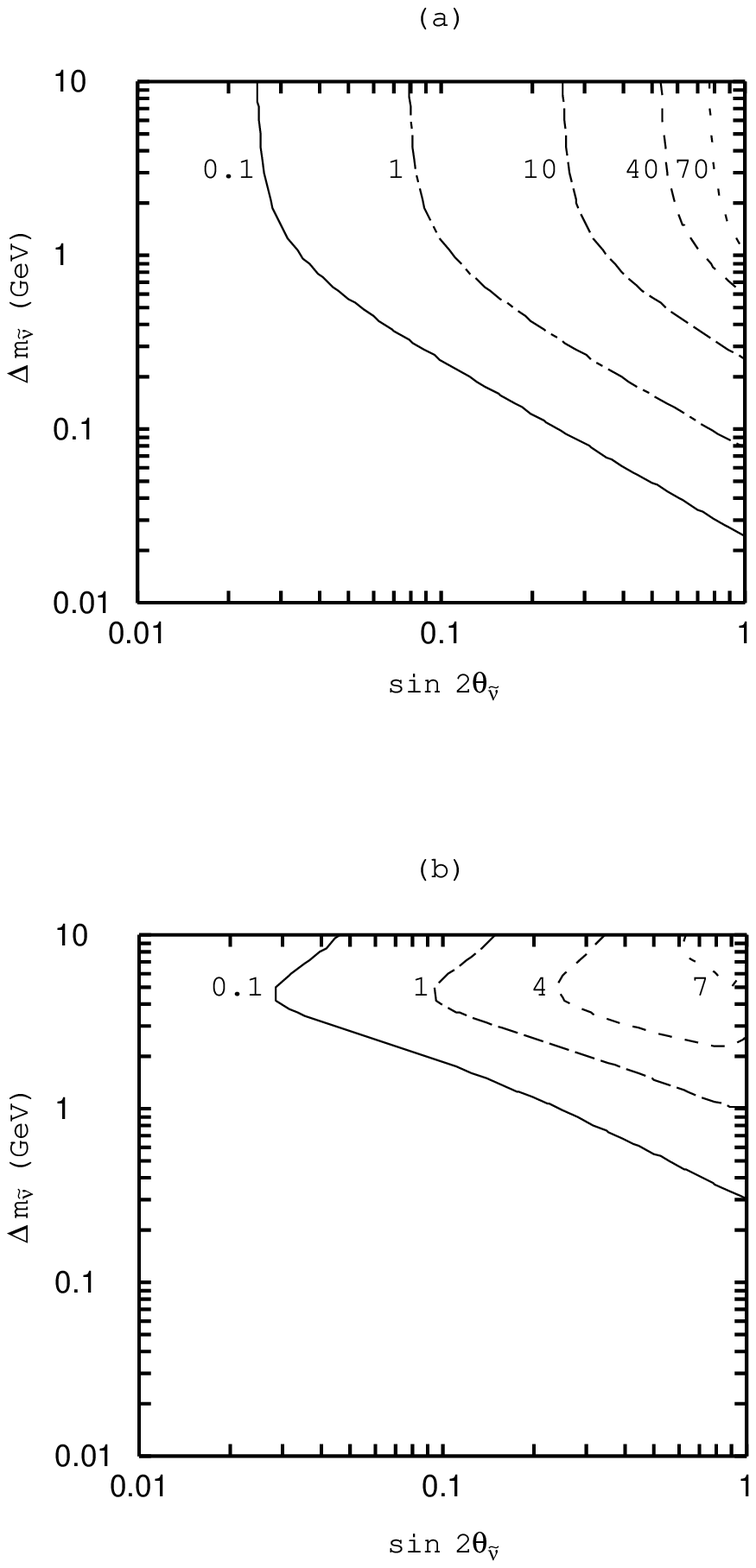,height=18cm}
\vspace*{1cm}
\caption{Contours of constant cross sections (in unit of fb) of 
 (a) $\epem \to \tau^+ e^- \charginoonep \charginoonem$ and 
 (b) $\epem \to \tau^+ e^- \neutralinotwo \neutralinotwo$ 
as functions of the mixing angle $\theta_{\snu}$ and 
the mass difference $\Delta m_{\snu}$ 
between the first and third generations
at the center-of-mass energy 500GeV.
We take $\tan\beta=10$ in the figures and other SUSY parameters 
are shown in TABLE I.}
\label{fig:sigmatanbeta10}
\end{center}
\end{figure}

Figures~\ref{fig:sigmatanbeta10} (a) and (b) are contours 
of constant cross sections of 
$\epem \to \sneu\sneu^c \to \tau^+ e^- \charginoonep \charginoonem$
and 
$\epem \to \slepL^+ \slepL^- \to \tau^+ e^- \neutralinotwo\neutralinotwo$
in the case of $\tan\beta=10$, respectively. In these figures 
we also fix the electron sneutrino mass as 180GeV, 
$m_{\charginoonepm}=100$ GeV, $\tan\beta=10$,
and other supersymmetric parameters are determined by solving 
the RGE's of the MSSM, similarly to the previous $\tan\beta=3$ case.

In the Figure~\ref{fig:sigmatanbeta10} (b) we can see 
nontrivial dependence of the cross section on $\Delta m_{\snu}$.
This is an effect of the left-right mixing elements 
of the third generation in the charged slepton mass-squared matrix, 
first pointed out in Ref.~\cite{HirouchiTanaka}.
Namely, in the charged slepton mass matrix among 
$\seL$, $\stauL$, and $\stauR$, 
\begin{eqnarray}
\left(               
\begin{array}{ccc}
 m^2_{L11} &  m^2_{L13}  &  0         \\
 m^2_{L31} &  m^2_{L33}  & m^2_{LR33} \\
     0     &  m^2_{LR33} & m^2_{R33}  
\end{array}
\right),
\end{eqnarray}
(Here we have ignored $m^2_{LR13}$.) the diagonalization of the 
lower-right $2\times2$ submatrix gives
\begin{eqnarray}
\left(               
\begin{array}{ccc}
 m^2_{L11}   &  m^2_{L13} \cos\theta_{\stau} 
                            & - m^2_{L13} \sin\theta_{\stau}   \\ 
   m^2_{L31} \cos\theta_{\stau} &  m^2_{\stau_2}     & 0       \\
 - m^2_{L31} \sin\theta_{\stau} &  0                 &  m^2_{\stau_1}
\end{array}
\right)
\end{eqnarray}
with
\begin{eqnarray}
m_{\tilde{\tau}_{1/2}}^2 
&=& 
\frac12 \left( m_{L33}^2 + m_{R33}^2 
\mp 
\sqrt{(m_{L33}^2 -m_{R33}^2)^2+ 4  (m_{LR33}^2)^2}
\right),
\nonumber \\
\tan2\theta_{\tilde{\tau}} &=& \frac{2m_{LR33}^2}{m_{L33}^2 -m_{R33}^2}.
\end{eqnarray}
In the limit of $m_{LR33} \to 0$ the difference 
between $m^2_{L11}$ and $m^2_{L33}$ is simply $(\Delta m_{\snu})^2$.
With presence of finite $m_{LR33}^2$, however, the stau masses are
given by 
\begin{eqnarray}
m_{\tilde{\tau}_{1/2}}^2 
&=& 
m_{R33/L33}^2 \mp \frac{(m_{LR33}^2)^2}{m_{L33}^2 -m_{R33}^2}
\end{eqnarray}
and $\stau_2$ becomes more degenerate in the mass with $\seL$
and as a consequence the mixing between $\seL$ and $\stau_2$ 
is enhanced.
This enhancement is maximized when 
$m^2_{L11} - m^2_{L33} \simeq \frac{(m_{LR33}^2)^2}{m_{L33}^2 -m_{R33}^2}$
is satisfied and this corresponds to $\Delta m_{\snu} \simeq 0.4$GeV
for our sample parameter set of $\tan\beta=10$.

In the Figures~\ref{fig:sigmatanbeta03} and \ref{fig:sigmatanbeta10}
we can see that the cross sections of 
$\epem \to \sneu\sneu^c \to \tau^+ e^- \charginoonep \charginoonem$ 
are several times as large as those of  
$\epem \to \slepL^+ \slepL^- \to \tau^+ e^- \neutralinotwo\neutralinotwo$.
This comes from the fact that in our sample parameter
destructive interference between $s$- and $t$-channel 
occurs in charged slepton pair production, while 
in sneutrino pair production they constructively contribute 
to the cross section~\cite{HNST}.

In the following subsections we will discuss the BG's for this 
process, and determine the reach of LC's.

\subsection{Backgrounds}
Now let us discuss the BG's for the signal (\ref{4jetsignal}).
Our situation here is almost parallel to that in Ref.~\cite{HNST}, where
the signal to be discovered is (\ref{4jetsignalHNST}).
In their work BG's and suitable cuts for reducing them are
discussed in detail and the results are almost straightforwardly 
applicable to our case.
Here we follow an outline of their discussion and 
apply their results to our case.

The main SM BG of the process (\ref{4jetsignal}) is from $ZW^+W^-$
production, and the total cross section of 
$\epem \to ZW^+W^-$ is about 39fb at $\sqrt{s}=500$GeV. 
This reduces to about 15fb~\cite{MiyamotoAPPI95}
if we require central production of gauge bosons $|\cos\theta_V|<0.8$
$(V=W^\pm, Z)$.
Then the cross section is 
\begin{eqnarray}
\sigma(\epem \to Z(\to \tau\tau \to \tau e )W^+W^-) = 0.17 {\rm fb},
\label{13SMBG}
\end{eqnarray}
and this is negligible compared to that of SUSY origin, as discussed
below.

Main SUSY BG is from tau lepton pair production
that results from slepton decay, and successive decay
of one of the two tau leptons to an electron, 
\begin{eqnarray}
\epem &\to& \stauL^+ \stauL^-  ~~~{\rm or}~~~  \sneutau\sneutau^c \nonumber\\
      &\to& \tau^\pm (\to e^\pm) \tau^\mp + 4jets + \E.   \label{4jetBG}
\end{eqnarray}
The tau pair production cross section 
$\sigma( \epem  \to \stauL^+ \stauL^-  ({\rm or} ~ \sneutau\sneutau^c) 
 \to \tau^+ \tau^- + 4jets + \E )$
is 2.5 fb for our sample parameter set of $\tan\beta=3$, 
and this BG can be reduced by applying suitable cuts
on the impact parameter $\sigma_{IP}$ and the electron energy $E_e$.

The cut on impact parameter (IP) will be effective in this case, 
since in (\ref{4jetBG}) the tau lepton decays after running
some finite length which is determined from the life time $\tau_\tau$
of tau lepton, 
which is typically of order of $c\tau_\tau \simeq 90\mu{\rm m}$.
Hence the more precisely we distinguish the distance 
between the track of the electron
and the interaction point, the more effectively 
we can tell the BG (\ref{4jetBG}) from signal candidates.
The distribution of IP from tau decay in flight is studied in 
Appendix E of Ref.~\cite{HNST} in case of $\tautomununu$ decay, and
their result is of course directly applicable to our $\tautoenunu$ case.
Recent detector designs of the JLC~\cite{JLC-I} and the NLC~\cite{NLC} 
show that the estimated values 
of the impact parameter resolution $\delta \sigma$ are given by
$\delta \sigma^2 = 11.4^2 + (28.8 /p\beta)^2 / \sin^3 \theta 
(\mu {\rm m}^2)$ and 
$\delta \sigma^2 = 2.6^2  + (13.7 /p\beta)^2 / \sin^3 \theta 
(\mu {\rm m}^2)$, respectively,
and an IP cut $\sigma_{IP}^{\rm cut}$ of order of 10 $\mu$m 
may be realistic.

The cut on the electron energy $E_e$ works as follows. 
The energy distribution of electrons 
which are produced by the tau lepton decay 
(namely, the energy distribution of BG electron in our case)
is different from that of electrons which
directly result from slepton decay
(that is, the energy distribution of signal electron).
Since the latter is two body decay, 
$\snu \to e^- \charginoonep$ and  $\snu^c \to e^+ \charginoonem$,
the energy distribution of electron is flat 
between kinematically determined energies 
$E_e^{\rm min}$ and $E_e^{\rm max}$.
Explicitly, $E_e^{\rm max}$ and $E_e^{\rm min}$ are given by 
\begin{eqnarray}
E_e^{\rm max(min)} = E_e^{C.M.} ( 1 \pm \beta_{\snu} \gamma_{\snu} )
\end{eqnarray}
with 
\begin{eqnarray}
\beta^2_{\snu}  &=& 1 - 4 m^2_{\snu} / s \\
\gamma^2_{\snu} &=& 1/(1-\beta^2_{\snu})
\end{eqnarray}
and
\begin{eqnarray}
E_e^{C.M.} = \frac{m^2_{\snu} - m^2_{\charginoone}}{2m_{\snu}} .
\end{eqnarray}
For our sample parameter set of $\tan\beta=3$, 
$E_e^{\rm min}=26.5$GeV and $E_e^{\rm max}=146$GeV.
On the other hand, the energy of electron
produced by the tau lepton decay is typically softer 
than that of signal electron in this case.
(The energy distribution of muon in tau decay in flight
is studied in Appendix E of Ref.~\cite{HNST},
and their results are directly applicable to our case
with replacement of the muon with an electron.)
Therefore if we discard event candidates in which electron energy
is smaller than certain energy 
we can reduce the BG and enhance the $S/N$ ratio.

In Ref.~\cite{HNST}, with the IP cut $\sigma_{IP}^{\rm cut}$
of 10$\mu$m and the muon energy cut $E > E_{\mu}^{\rm min}$
they obtain the muon misidentification rate $p_\mu=0.02$ 
and the tau identification rate $p_\tau=0.88$, respectively.
We apply their result of $\mu\tau$ case to our $e\tau$ case, 
and take the electron misidentification rate $p_e=0.02$ and
the tau identification rate $p_\tau=0.88$, respectively.
These are reasonable choices since we have chosen 
our sample parameters almost the same as theirs.

In addition to these cuts we will have to apply isolation cuts
for the leptons and other cuts to select signal events.
We take into account them as an overall factor $A$ common to 
$\tau\tau$ and $e\tau$ events. 
We assume $A=0.3$ as done in Ref.~\cite{HNST} in order to 
estimate the collider reach. 

The signal and the BG cross sections after applying these cuts are
\begin{eqnarray}
\sigma_{sig}= A p_{\tau} 
  (  \sigma(\epem \to \tau^+ e^- 4jets \E) 
   + \sigma(\epem \to \tau^- e^+ 4jets \E)   )
\end{eqnarray}
and 
\begin{eqnarray}
\sigma_{bg}= 2 A p_{e} p_{\tau} \sigma(\epem \to \tau^+ \tau^- 4jets \E)
\end{eqnarray}
respectively.

\subsection{Collider Reach}
\label{sec:collider-reach}

Now we discuss possible reach of mass difference $\Delta m_{\snu}$ and
mixing angle $\theta_{\snu}$ at LC's.

\begin{figure}
\begin{center}
\leavevmode
\psfig{file=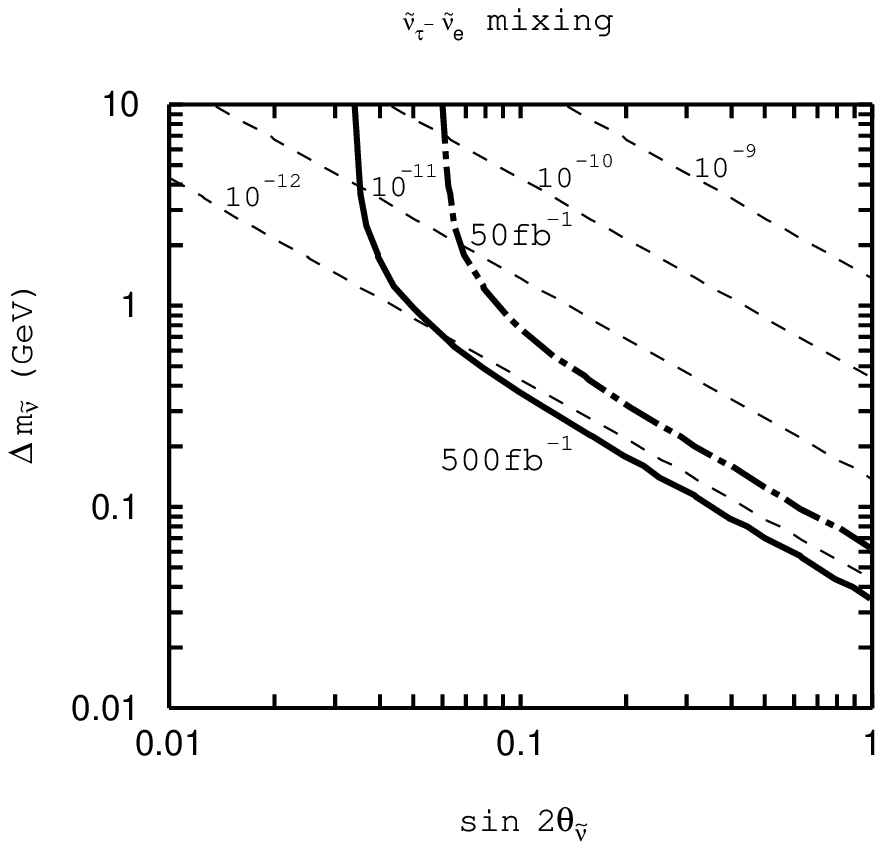,height=10cm}
\vspace*{1cm}
\caption{
Contours of $5\sigma$ discovery reaches 
by LC's with integrated luminosities 
${\cal L}=50  {\rm fb}^{-1}$ (dashed-dot contour) and 
${\cal L}=500 {\rm fb}^{-1}$ (solid contour)
as functions of the mixing angle $\theta_{\snu}$ and 
the mass difference $\Delta m_{\snu}$ 
between the first and third generations. 
Contours for constant branching ratios of $\tau \to e \gamma$
are also shown, and each dashed line means 
the contour on which the branching ratio of $\tau \to e \gamma$ 
is $10^{-9}, 10^{-10}, 10^{-11}$, and $10^{-12}$, respectively.
We take $\tan\beta=3$ in the figure 
and other SUSY parameters are shown in TABLE I.}
\label{fig:5sigmareach}
\end{center}
\end{figure}

In Figure~\ref{fig:5sigmareach} we show 
contours of $5\sigma$ reach of future LC's for
integrated luminosity ${\cal L}= 50{\rm fb}^{-1}$ (the dashed-dot contour) 
and ${\cal L}= 500{\rm fb}^{-1}$ (the solid contour).
In the figure, we take our sample parameter set of $\tan\beta=3$,
tabulated in Table I. We have used the values of decay branching ratios of
SUSY particles (sleptons, the lighter chargino, 
and the second lightest neutralino)
calculated in Ref.~\cite{HNST}. 
(They are listed in Table II and III.)

From the figure we can see that for our sample parameters
with integrated luminosity 
${\cal L}=50 {\rm fb}^{-1}(500 {\rm fb}^{-1})$ it is possible to reach 
mixing angle $\sin 2\theta_{\tilde{\nu}} \gtrsim 0.06 (0.04)$ and 
mass difference  $\Delta m_{\tilde{\nu}} \gtrsim 0.07 (0.04)$ GeV 
for sneutrinos in the first and third generations 
at the statistical significance of 5$\sigma$.

Contours for constant branching ratios of $\tautoeg$ are also
plotted in the figure. 
We can see that at least for our sample parameter set
there are some parameter region in which 
future LC's can probe the slepton flavor mixing 
between the first and third generations
beyond the reach of $\tautoeg$ 
by several orders of magnitude in sensitivity.

For our sample parameter set of $\tan\beta=10$   
we can expect almost the same reach 
as that for our $\tan\beta=3$ parameter set.
This is because left-handed slepton (especially sneutrino) 
pair production cross sections and 
decay branching ratio of $\charginoone\to 2 jets \LSP$
are almost the same for both two cases, 
as we can see from Figures 1, 2, and Table II.

\section{Application to $\snu_\e$-$\snu_\mu$ mixing}
\label{sec:12-mixing}

So far in this article we have assumed that 
the flavor mixing between the first and third generations
is the only LFV mixing in the left-handed slepton mass matrix.
If there is left-handed slepton flavor mixing 
between the first and second generations
it contributes to the $\mutoeg$ decay rate, 
which is severely constrained by 
experiments and further sensitive experiment is
being prepared~\cite{PSIprop}. 
In the following we estimate the reach of LC's in probing
the left-handed slepton flavor mixing
between the first and second generations.

Here we adopt 
\begin{eqnarray}
\epem &\to& \slepL^+ \slepL^-  ~~~{\rm or}~~~  \sneu\sneu^c   \nonumber  \\
      &\to& e^\pm \mu^\mp + 4jets + \E      \label{4jetsignalme}
\end{eqnarray}
as LFV signal. For our sample parameter set of $\tan\beta=3$
the cross section of this signal is as large as
\begin{eqnarray}
\sigma(\epem \to e^+ \mu^- + 4jets + \E) 
& \simeq &
4 \chi_{\snu_{12}}  \sin^2 2 \theta_{\snu_{12}}
( 1 -  \chi_{\snu_{12}} \sin^2 2 \theta_{\snu_{12}} ) 
 \times 30  ~{\rm fb}  ,
\label{sigma-4jetsignalemu} 
\end{eqnarray}
since also in this case $t$-channel pair production of 
the left-handed sleptons of the first generation is available.
Here $\chi_{\snu_{ij}}$  $(i,j=1,...,3)$ are
\begin{eqnarray}
\chi_{\snu_{ij}} &\equiv& 
\frac{ (x_{ij}^{(\snu)})^2 }{ 2 ( 1 + (x_{ij}^{(\snu)})^2 )} 
\label{defofchi-ij}
\end{eqnarray}
with
\begin{eqnarray}
x_{ij}^{(\snu)} &\equiv& 
\frac{ m^2_{\snu_i} - m^2_{\snu_j} }{ 2 [ m \Gamma ]_{ij} } 
\end{eqnarray}
and
\begin{eqnarray}
[ m \Gamma ]_{ij} & \equiv & 
( m_{\snu_i}   \Gamma_{\snu_i} + m_{\snu_j} \Gamma_{\snu_j} )/2 .
\end{eqnarray}
($\snu_1$, $\snu_2$, and $\snu_3$ stand for
$\snu_e$, $\snu_\mu$, and $\snu_\tau$, respectively. 
$m_{\snu_i}$ $(i=1,...,3)$ is the mass of $\snu_i$ 
and $\Gamma_{\snu_i}$ the decay width.)
And $\theta_{\snu_{ij}}$ $(i \neq j)$ are defined by
\begin{eqnarray}
\tan 2 \theta_{\snu_{ij}} &\equiv& 
\frac{2 m^2_{\snu ij}}
{m^2_{\snu ii} - m^2_{\snu jj}} ~ ,
~~~~ ({\rm no \ sum \ over \ } i {\rm \  and \ } j)
\end{eqnarray}
where $m^2_{\snu ij}$ is the $(i,j)$ element of the mass-squared matrix 
of sneutrinos.

The main SM BG for the signal (\ref{4jetsignalme}) is from 
$\epem \to Z(\to \tau\tau \to \mu e )W^+W^-$, similarly to 
13-mixing case eq.~(\ref{13SMBG}), 
and with the requirement of central production 
of gauge bosons $|\cos\theta_V|<0.8$ $(V=W^\pm, Z)$
the cross section is as small as
\begin{eqnarray}
\sigma(\epem \to Z(\to \tau\tau \to \mu e )W^+W^-) = 0.031 {\rm fb}
\label{12SMBG}
\end{eqnarray}
because of the suppressions by 
${\rm Br}(\tau^- \to e^-   \bar{\nu}_e   \nu_\tau) = 0.18$~\cite{PDG00} and
${\rm Br}(\tau^- \to \mu^- \bar{\nu}_\mu \nu_\tau) = 0.17$~\cite{PDG00}.
On the other hand, there are two types of SUSY BG. 
One is slepton-mixing parameter independent and the other dependent.
The former is 
\begin{eqnarray}
\epem &\to& \stauL^+ \stauL^-  ~~~{\rm or}~~~  \sneutau\sneutau^c \nonumber\\
      &\to& \tau^\pm (\to e^\pm) \tau^\mp (\to \mu^\mp) 
            + 4jets + \E 
\label{4jetBGtaumutaue} 
\end{eqnarray}
and the cross section is 
$\sigma(\epem \to \tau^\pm (\to e^\pm) \tau^\mp (\to \mu^\mp) 
4jets \E ) = 0.15$fb. This can be further reduced by applying 
IP cuts and energy cuts on electron and muon, 
as done in 13-case in this article.
The latter BG processes are induced by slepton oscillations due to
23-mixing and 13-mixing, namely,
\begin{eqnarray}
\epem &\to& \slepL^+ \slepL^-  ~~~{\rm or}~~~  \sneu\sneu^c \nonumber\\
      &\to& \mu^\pm \tau^\mp (\to e^\mp) + 4jets + \E   
\label{4jetBGmutaue}
\end{eqnarray}
and
\begin{eqnarray}
\epem &\to& \slepL^+ \slepL^-  ~~~{\rm or}~~~  \sneu\sneu^c \nonumber\\
      &\to& e^\pm \tau^\mp (\to \mu^\mp) + 4jets + \E ,  
\label{4jetBGetaumu}
\end{eqnarray}
respectively, and their cross sections are 
\begin{eqnarray}
\sigma(\epem \to \mu^\pm \tau^\mp (\to e^\mp) + 4jets + \E) 
& \simeq &
2 \chi_{\snu_{23}} ( 3 - 4 \chi_{\snu_{23}} ) 
\sin^2 2 \theta_{\snu_{23}} \times 0.43  ~{\rm fb} 
\label{sigma-4jetBGmutaue} 
\end{eqnarray}
and 
\begin{eqnarray}
\sigma(\epem \to e^\pm \tau^\mp (\to \mu^\mp) + 4jets + \E) 
& \simeq &
4 \chi_{\snu_{13}}  \sin^2 2 \theta_{\snu_{13}}
( 1 -  \chi_{\snu_{13}} \sin^2 2 \theta_{\snu_{13}} ) 
 \times 10.4  ~{\rm fb}  
\label{sigma-4jetBGetaumu} 
\end{eqnarray}
for our sample parameter set of $\tan\beta=3$.
Therefore the processes (\ref{4jetBGmutaue}) and (\ref{4jetBGetaumu})
can be dominant BG, depending on the parameters.
However, hereafter we neglect the BG's 
(\ref{4jetBGmutaue}) and (\ref{4jetBGetaumu}) and assume 
(\ref{4jetBGtaumutaue}) to be the dominant BG, in order to estimate
the collider reach.

Applying the same energy cuts and the same IP cuts on electron and muon
as done in the previous 13-mixing case, 
the signal and BG cross sections are 
\begin{eqnarray}
\sigma_{sig}= A
  (  \sigma(\epem \to \mu^+ e^- 4jets \E) 
   + \sigma(\epem \to \mu^- e^+ 4jets \E)   )
\end{eqnarray}
and 
\begin{eqnarray}
\sigma_{bg}= 2 A p_{e} p_{\mu} \sigma(\epem \to \tau^+ \tau^- 4jets \E), 
\end{eqnarray}
where $A$ is the overall factor that parameterize the effect of 
other cuts to isolate leptons
and $p_e(p_\mu)$ is misidentification rate of electron (muon), 
and we assume $A=0.3$ and $p_e = p_\mu = 0.02$ in order to estimate 
collider reach, as done in the 13-mixing case.

\begin{figure}
\begin{center}
\leavevmode
\psfig{file=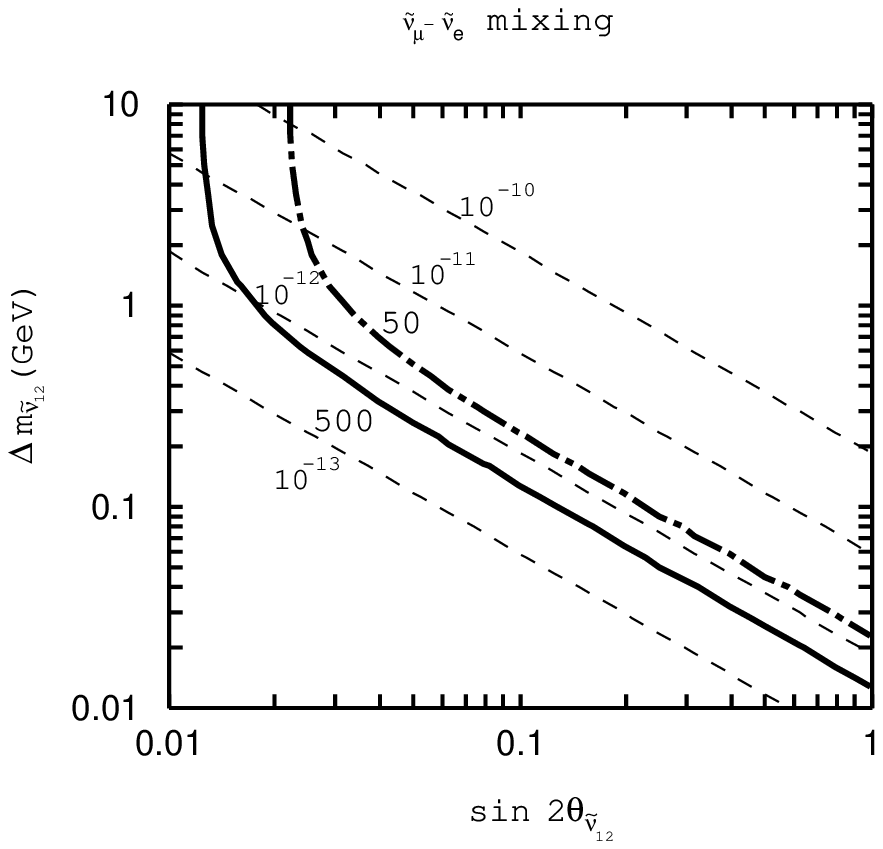,height=10cm}
\vspace*{1cm}
\caption{
Contours of $5\sigma$ discovery reaches 
by LC's with integrated luminosities 
${\cal L}=50  {\rm fb}^{-1}$ (dashed-dot contour) and 
${\cal L}=500 {\rm fb}^{-1}$ (solid contour)
as functions of the mixing angle $\theta_{\snu_{12}}$ and 
the mass difference $\Delta m_{\snu_{12}}$ 
between the first and second generations. 
Contours for constant branching ratios of $\mu \to e \gamma$
are also shown, and each dashed line means 
the contour on which the branching ratio of $\mu \to e \gamma$ 
is $10^{-10}, 10^{-11}, 10^{-12}$, and $10^{-13}$, respectively.
We take $\tan\beta=3$ in the figure 
and other SUSY parameters are shown in TABLE I.}
\label{fig:5sigmareachme}
\end{center}
\end{figure}

$5\sigma$ reach in 12-mixing is shown in Figure~\ref{fig:5sigmareachme}.
In the figure we have taken the parameter set of $\tan\beta=3$
shown in Table I.
From the figure we can see that with $\epem$ collider with
${\cal L}=50 {\rm fb}^{-1}(500 {\rm fb}^{-1})$ it may be possible to reach 
mixing angle $\sin 2\theta_{\tilde{\nu}_{12}} \gtrsim 0.03 (0.02)$ and 
mass difference $\Delta m_{\tilde{\nu}_{12}}  \gtrsim 0.03 (0.02)$ GeV 
for sneutrinos in the first and second generations 
at the statistical significance of 5$\sigma$.
Contours of constant values of Br$(\mutoeg)$ are also plotted 
in the figure. 
The reaches of LC's correspond to Br$(\mutoeg) \simeq 10^{-12}$.
Since planned $\mutoeg$ search~\cite{PSIprop} will probe Br$(\mutoeg)$ 
down to the sensitivity of $10^{-14}$, 
left-handed slepton oscillation search is not so powerful as 
$\mutoeg$ search at least for our sample parameter set.

\section{Application to radiatively induced sneutrino mixings
via right-handed neutrino Yukawa couplings}
\label{sec:radiativemixing}

In the above analysis we have not specified any origin of 
the LFV off-diagonal elements of the slepton mass matrix.
In this section we extend our analysis 
and discuss possible reach of LC's in probing 13-mixing 
in the case where 
sizable 23-mixing and non-vanishing 12- and 13-mixings 
in the left-handed slepton mass matrix
are radiatively induced via right-handed neutrino 
Yukawa couplings.

Introduction of the right-handed neutrino is one of the most
attractive extensions of the MSSM, since it naturally explains
the tiny neutrino masses via the seesaw mechanism~\cite{seesaw}.
Here let us consider the case in which 
there exist three generations of 
the right-handed neutrino superfields $\overline{N}$.
In this case, the superpotential $W$ of 
the Higgs sector and the lepton sector is given as
\begin{eqnarray}
W &=&  f_{\nu_{ij}} H_2 \overline{N}_i L_j 
     + f_{e_{ij}}   H_1 \overline{E}_i L_j 
     + \frac12 M_{\nu_i\nu_j} \overline{N}_i \overline{N}_j 
     + \mu H_1 H_2.
\label{MSSMRN}
\end{eqnarray}
Here $H_1$ and $H_2$ are chiral superfields 
of the Higgs doublets in the MSSM. 
$L_i$ and $\overline{E}_i$ are chiral superfields 
of the lepton doublets and the right-handed charged leptons, 
respectively.
$i$ and $j$ are generation indices and run from 1 through 3.  
After suitable redefinition of the fields, 
the Yukawa coupling constants and the Majorana masses 
can be taken as 
\begin{eqnarray}
f_{\nu_{ij}} &=& f_{\nu_i} V_{D ij},\nonumber\\
f_{e_{ij}}   &=& f_{e_i}\delta_{ij},       \label{redefYukawa} \\
M_{\nu_i\nu_j} &=& U_{ik}^{\ast} M_{\nu_{k}} U^\dagger_{kj},   \nonumber
\end{eqnarray}
where $V_D$ and $U$ are unitary matrices. 
By integrating out the right-handed neutrinos
which are supposed superheavy, 
we obtain the mass matrix $(m_{\nu})$ of the left-handed neutrinos 
\begin{eqnarray}
(m_{\nu})_{ij} &=& 
V_{D ik}^\top (\overline{m}_{\nu})_{kl} V_{D lj},
\end{eqnarray}
where the matrix $\overline{m}_{\nu}$ is 
\begin{eqnarray}
\overline{m}_{\nu} &=& 
\left(
\begin{array}{ccc}
m_{{\nu_1}D} & & \\
& m_{{\nu_2}D} & \\
& & m_{{\nu_3}D}
\end{array}
\right)
U^\top
\left(
\begin{array}{ccc}
\frac1{M_{\nu_1}} & & \\
& \frac1{M_{\nu_2}} & \\
& & \frac1{M_{\nu_3}}
\end{array}
\right)
U
\left(
\begin{array}{ccc}
m_{{\nu_1}D} & & \\
& m_{{\nu_2}D} & \\
& & m_{{\nu_3}D}
\end{array}
\right)                \nonumber\\
&\equiv&
V^\top_M 
\left(
\begin{array}{ccc}
m_{{\nu_e}} & & \\
& m_{{\nu_\mu}} & \\
& & m_{{\nu_\tau}}
\end{array}
\right)
V_M .
\end{eqnarray}
Here, $m_{{\nu_i}D}=f_{\nu_i}v\sin\beta/\sqrt{2}$
and $V_M$ is a unitary matrix.\footnote{ 
$\langle h_1 \rangle=(v\cos\beta/\sqrt{2},0)^\top$ and
$\langle h_2 \rangle=(0,v\sin\beta/\sqrt{2})^\top$ with $v\simeq 246$GeV.}
We can see that the mixings among left-handed neutrinos are determined from
$V_D$ and $U$.
Let us assume hereafter that $U$ is a unit matrix for simplicity. 
Then the mixing matrix among left-handed neutrinos is $V_D$ itself.
Under the assumption that $U={\bf 1}$, 
if the atmospheric neutrino oscillation
is between $\nu_\mu$ and $\nu_\tau$,
we can expect that $V_{D 32}$ is of order of 1. 
If we further assume hierarchy among neutrino masses,
$m_{\nu_e} \ll m_{\nu_\mu} \ll m_{\nu_\tau}$, 
the heaviest neutrino mass $m_{\nu_\tau}$ is related with 
the third generation right-handed Majorana mass $M_{\nu_3}$
and the third generation neutrino Yukawa coupling constant $f_{\nu_3}$ as
\begin{eqnarray}
m_{\nu_\tau} = \frac{f^2_{\nu_3} v^2 \sin^2\beta}{M_{\nu_3}} .
\label{eq:seesaw}
\end{eqnarray}
In the following discussion we use eq.~(\ref{eq:seesaw})
in estimating rates of LFV processes.

Now let us discuss the LFV off-diagonal elements
of the left-handed slepton mass matrix 
which are radiatively induced by LFV Yukawa coupling constant matrix
in eqs.~(\ref{redefYukawa}).
In the equations we saw that 
we cannot diagonalize the Yukawa coupling constant matrices
$f_{\nu_{ij}}$ and $f_{e_{ij}}$ simultaneously.
In the minimal SUGRA scenario~\cite{Nilles}, 
these non-diagonal elements
of the Yukawa coupling constant matrix generate
the off-diagonal elements of the slepton mass matrix
via radiative corrections~\cite{LFVneutrino}.
In this scenario we assume the slepton soft mass matrices to be 
generation-diagonal at the gravitational scale $\Mgrav$ as
\begin{eqnarray}
&(m_{\tilde L}^2)_{ij}=(m_{\tilde e}^2)_{ij}= m_0^2  \delta_{ij} , &
\nonumber\\
&{\tilde m}_{h1}^2 = {\tilde m}_{h2}^2 = m_0^2 , &    
\label{mSUGRA} \\
& A_{e_{ij}}=f_{e_{ij}}a_0, \,  A_{\nu_{ij}}=f_{\nu_{ij}}a_0. &
\nonumber
\end{eqnarray}
Here $(m_{\tilde L}^2)_{ij}$ and $(m_{\tilde e}^2)_{ij}$ are
the $(i,j)$ element of the soft SUSY breaking mass-squared matrices 
of the left-handed and the right-handed sleptons, respectively. 
${\tilde m}_{h1}$ and ${\tilde m}_{h2}$ are
the soft masses for the Higgs bosons,
$A_{e_{ij}}$ is the soft trilinear scalar
coupling constant among the Higgs boson, 
the $i$-th generation right-handed charged slepton 
and the $j$-th generation left-handed slepton 
and $A_{\nu_{ij}}$ that among
the Higgs boson, the $i$-th generation right-handed sneutrino 
and the $j$-th generation left-handed slepton.
Under the assumptions of eqs.~(\ref{mSUGRA}) 
there is no LFV at the tree level at $\Mgrav$.
However, these relations are subject to radiative corrections, and
at the weak scale they are approximately given as 
\begin{eqnarray}
(m_{\tilde L}^2)_{ij} &\simeq& 
-\frac1{8\pi^2} (3m_0^2+a_0^2) V_{D ki}^\ast V_{D kj} f_{\nu_k}^2
\log \frac{M_{\rm grav}}{M_{\nu_k}},
\label{generatedLFV} \\
(m^2_{\tilde e})_{ij} &\simeq&   0,       \\
A_e^{ij} &\simeq& 
-\frac{3}{8\pi^2} a_0 f_{e_i} V_{D ki}^\ast V_{D kj} f_{\nu_k}^2 
\log \frac{M_{\rm grav}}{M_{\nu_k}}          
\end{eqnarray}
for $i\neq j$.
These parameters induce rare LFV processes such as 
$\muegamma, \tauegamma$, and so on.
Especially from large $V_{D32}$
we can expect sizable $(m_{\tilde L}^2)_{32}$ 
and therefore sizable $\taumugamma$ rate.
Comparison between possible reach of the future lepton colliders in 
$(m_{\tilde L}^2)_{32}$ 
and that of $\taumugamma$ search is discussed in Ref.~\cite{HNST}.
Here we extend the analysis to the case 
in which $V_{D31}$ is also non-vanishing.

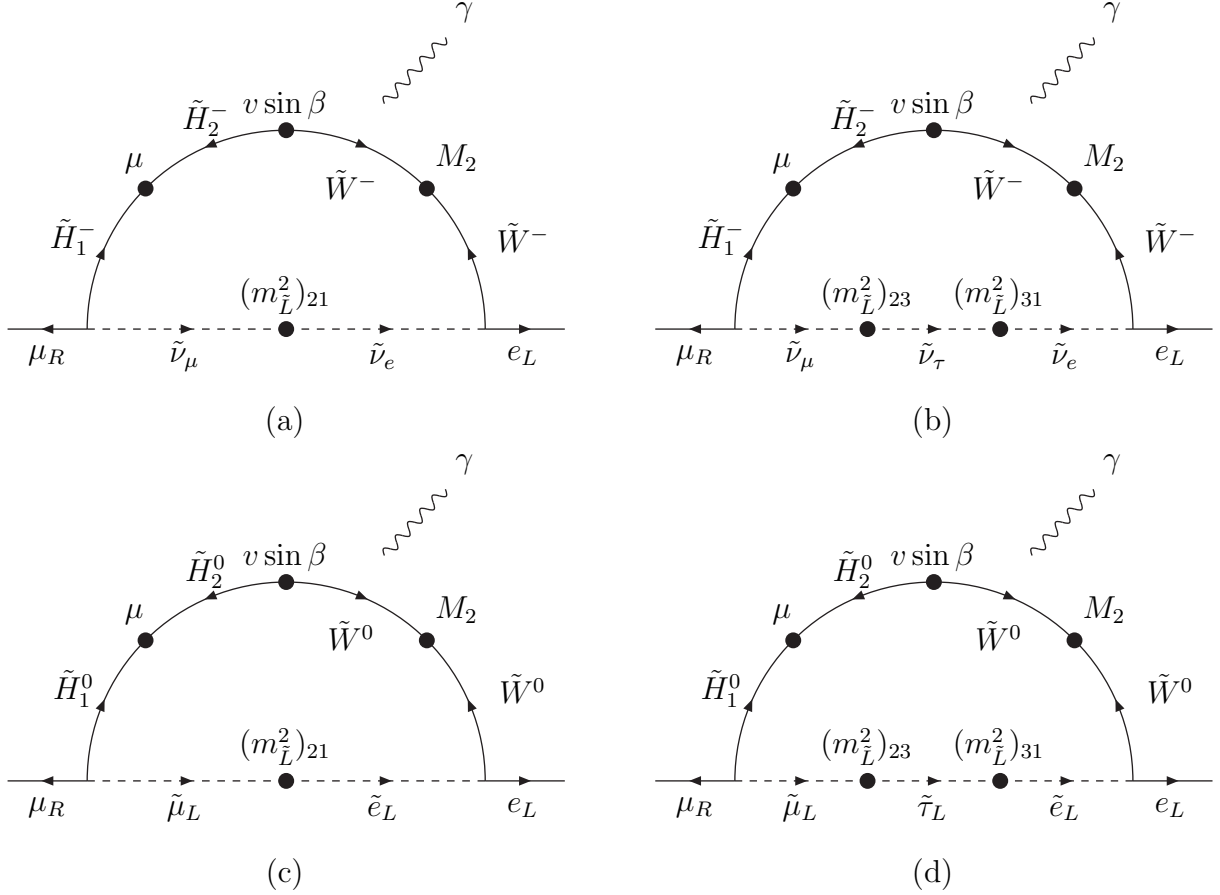
\begin{figure}
\begin{center} 
\vspace*{3cm}
\begin{picture}(455,140)(30,-20)
%%%%%%%%%%%%%%%%%%%%%%%%%%%%%%%%%%%%%%%%% graph (a) %%%%%%%%
\ArrowArcn(135,25)(75,180,135)
\ArrowArc(135,25)(75,90,135)
\ArrowArcn(135,25)(75,90,45)
\ArrowArc(135,25)(75,0,45)
\Vertex(188,78){3}
\Vertex(82,78){3}
\Vertex(135,100){3}
\Text(135,110)[]{$v\sin\beta$}
\Text(55,60)[]{$\tilde{H}_1^-$}
\Text(78,88)[]{$\mu$}
\Text(105,105)[]{$\tilde{H}_2^-$}
\Text(160,80)[]{$\tilde{W}^-$}
\Text(200,90)[]{$M_2$}
\Text(225,60)[]{$\tilde{W}^-$}

\ArrowLine(60,25)(30,25)
\DashArrowLine(60,25)(135,25){3}             \Vertex(135,25){3}
\DashArrowLine(135,25)(210,25){3}        
\ArrowLine(210,25)(240,25)

\Text(45,15)[]{$\mu_R$}
\Text(97,15)[]{$\tilde{\nu}_{\mu}$}
\Text(172,15)[]{$\tilde{\nu}_e$}
\Text(225,15)[]{$e_L$}

\Text(135,38)[]{$(m^2_{\tilde{L}})_{21}$}

\Photon(172,110)(195,135){2}{5}
\Text(203,145)[]{$\gamma$}

\Text(135,-10)[]{(a)}
%%%%%%%%%%%%%%%%%%%%%%%%%%%%%%%%%%%%%%%%% graph (b) %%%%%%%%
%\SetOffset(260,0)
\SetOffset(245,0)
\ArrowArcn(135,25)(75,180,135)
\ArrowArc(135,25)(75,90,135)
\ArrowArcn(135,25)(75,90,45)
\ArrowArc(135,25)(75,0,45)
\Vertex(188,78){3}
\Vertex(82,78){3}
\Vertex(135,100){3}
\Text(135,110)[]{$v\sin\beta$}
\Text(55,60)[]{$\tilde{H}_1^-$}
\Text(78,88)[]{$\mu$}
\Text(105,105)[]{$\tilde{H}_2^-$}
\Text(160,80)[]{$\tilde{W}^-$}
\Text(200,90)[]{$M_2$}
\Text(225,60)[]{$\tilde{W}^-$}

\ArrowLine(60,25)(30,25)
\DashArrowLine(60,25)(110,25){3}             \Vertex(110,25){3}
\DashArrowLine(110,25)(160,25){3}            \Vertex(160,25){3}
\DashArrowLine(160,25)(210,25){3}        
\ArrowLine(210,25)(240,25)

\Text(45,15)[]{$\mu_R$}
\Text(85,15)[]{$\tilde{\nu}_{\mu}$}
\Text(135,15)[]{$\tilde{\nu}_{\tau}$}
\Text(185,15)[]{$\tilde{\nu}_e$}
\Text(225,15)[]{$e_L$}

\Text(110,38)[]{$(m^2_{\tilde{L}})_{23}$}
\Text(160,38)[]{$(m^2_{\tilde{L}})_{31}$}

\Photon(172,110)(195,135){2}{5}
\Text(203,145)[]{$\gamma$}

\Text(135,-10)[]{(b)}

\end{picture} 

%\begin{picture}(530,170)(40,-20)
\begin{picture}(455,170)(30,-20)
%%%%%%%%%%%%%%%%%%%%%%%%%%%%%%%%%%%%%%%%% graph (c) %%%%%%%%
\ArrowArcn(135,25)(75,180,135)
\ArrowArc(135,25)(75,90,135)
\ArrowArcn(135,25)(75,90,45)
\ArrowArc(135,25)(75,0,45)
\Vertex(188,78){3}
\Vertex(82,78){3}
\Vertex(135,100){3}
\Text(135,110)[]{$v\sin\beta$}
\Text(55,60)[]{$\tilde{H}_1^0$}
\Text(78,88)[]{$\mu$}
\Text(105,105)[]{$\tilde{H}_2^0$}
\Text(160,80)[]{$\tilde{W}^0$}
\Text(200,90)[]{$M_2$}
\Text(225,60)[]{$\tilde{W}^0$}

\ArrowLine(60,25)(30,25)
\DashArrowLine(60,25)(135,25){3}             \Vertex(135,25){3}
\DashArrowLine(135,25)(210,25){3}        
\ArrowLine(210,25)(240,25)

\Text(45,15)[]{$\mu_R$}
\Text(97,15)[]{$\tilde{\mu}_L$}
\Text(172,15)[]{$\tilde{e}_L$}
\Text(225,15)[]{$e_L$}

\Text(135,38)[]{$(m^2_{\tilde{L}})_{21}$}

\Photon(172,110)(195,135){2}{5}
\Text(203,145)[]{$\gamma$}

\Text(135,-10)[]{(c)}

%%%%%%%%%%%%%%%%%%%%%%%%%%%%%%%%%%%%%%%%% graph (d) %%%%%%%%
%\SetOffset(260,0)
\SetOffset(245,0)
\ArrowArcn(135,25)(75,180,135)
\ArrowArc(135,25)(75,90,135)
\ArrowArcn(135,25)(75,90,45)
\ArrowArc(135,25)(75,0,45)
\Vertex(188,78){3}
\Vertex(82,78){3}
\Vertex(135,100){3}
\Text(135,110)[]{$v\sin\beta$}
\Text(55,60)[]{$\tilde{H}_1^0$}
\Text(78,88)[]{$\mu$}
\Text(105,105)[]{$\tilde{H}_2^0$}
\Text(160,80)[]{$\tilde{W}^0$}
\Text(200,90)[]{$M_2$}
\Text(225,60)[]{$\tilde{W}^0$}

\ArrowLine(60,25)(30,25)
\DashArrowLine(60,25)(110,25){3}             \Vertex(110,25){3}
\DashArrowLine(110,25)(160,25){3}            \Vertex(160,25){3}
\DashArrowLine(160,25)(210,25){3}        
\ArrowLine(210,25)(240,25)

\Text(45,15)[]{$\mu_R$}
\Text(85,15)[]{$\tilde{\mu}_L$}
\Text(135,15)[]{$\tilde{\tau}_L$}
\Text(185,15)[]{$\tilde{e}_L$}
\Text(225,15)[]{$e_L$}

\Text(110,38)[]{$(m^2_{\tilde{L}})_{23}$}
\Text(160,38)[]{$(m^2_{\tilde{L}})_{31}$}

\Photon(172,110)(195,135){2}{5}
\Text(203,145)[]{$\gamma$}

\Text(135,-10)[]{(d)}

\end{picture} 

\vspace*{1cm}

\caption{Dominant Feynman diagrams in $\mu \to e \gamma$ decay
when $\tan\beta\gsim1$ and 
off-diagonal elements of the right-handed slepton soft mass matrix
are negligibly small.
In the diagrams, $(m^2_{\tilde{L}})_{ij}$ denotes the $(i,j)$ element of 
the left-handed slepton soft mass matrix.
$\tilde{\tau}_{L}$,  $\tilde{\mu}_{L}$ and $\tilde{e}_{L}$
are the left-handed stau, smuon, and selectron, respectively,
and $\tilde{\nu}_{\tau}$,  $\tilde{\nu}_{\mu}$ and $\tilde{\nu}_{e}$ 
the tau sneutrino, the mu sneutrino and the electron sneutrino,
respectively.
$\tilde{H}_{1}$ and $\tilde{H}_{2}$ are Higgsino,
and $\tilde{W}$  W-ino.
The symbol $\mu$ is the Higgsino mass. 
The arrows represent the chirality.}

\label{fig:muegammadiag}

\end{center}
\end{figure}

In the case where $V_{D32}$ is of order of 1 and
$V_{D31}$ is not zero, all the off-diagonal elements 
in $(m^2_{\tilde{L}})$ are radiatively induced.
They are approximately given by
\begin{eqnarray}
(m_{\tilde L}^2)_{21} &\simeq& 
-\frac1{8\pi^2} (3m_0^2+a_0^2) V_{D 32}^\ast V_{D 31} f_{\nu_3}^2
\log \frac{M_{\rm grav}}{M_{\nu_3}},     
\label{21element}    \\
(m_{\tilde L}^2)_{31} &\simeq& 
-\frac1{8\pi^2} (3m_0^2+a_0^2) V_{D 33}^\ast V_{D 31} f_{\nu_3}^2
\log \frac{M_{\rm grav}}{M_{\nu_3}},       
\label{31element}    \\
(m_{\tilde L}^2)_{32} &\simeq& 
-\frac1{8\pi^2} (3m_0^2+a_0^2) V_{D 33}^\ast V_{D 32} f_{\nu_3}^2
\log \frac{M_{\rm grav}}{M_{\nu_3}}.         
\label{32element}   
\end{eqnarray}
Here we have neglected $f_{\nu_2}$ and $f_{\nu_1}$.
From these off-diagonal elements, non-trivial $\muegamma$ rate is 
predicted via diagrams shown in Figure~\ref{fig:muegammadiag}, 
in addition to sizable $\taumugamma$ and 
non-vanishing $\tauegamma$ rates.
The left-handed slepton mixing parameters can be
most severely constrained 
from the experimental bound of $\muegamma$ rate. 
Then with sizable $(m^2_{\tilde{L}})_{32}$
the slepton oscillation as a probe of $(m_{\tilde L}^2)_{31}$  
is not only a competition with $\tauegamma$  
but also that with $\muegamma$.

\begin{figure}
\begin{center}
\leavevmode
\psfig{file=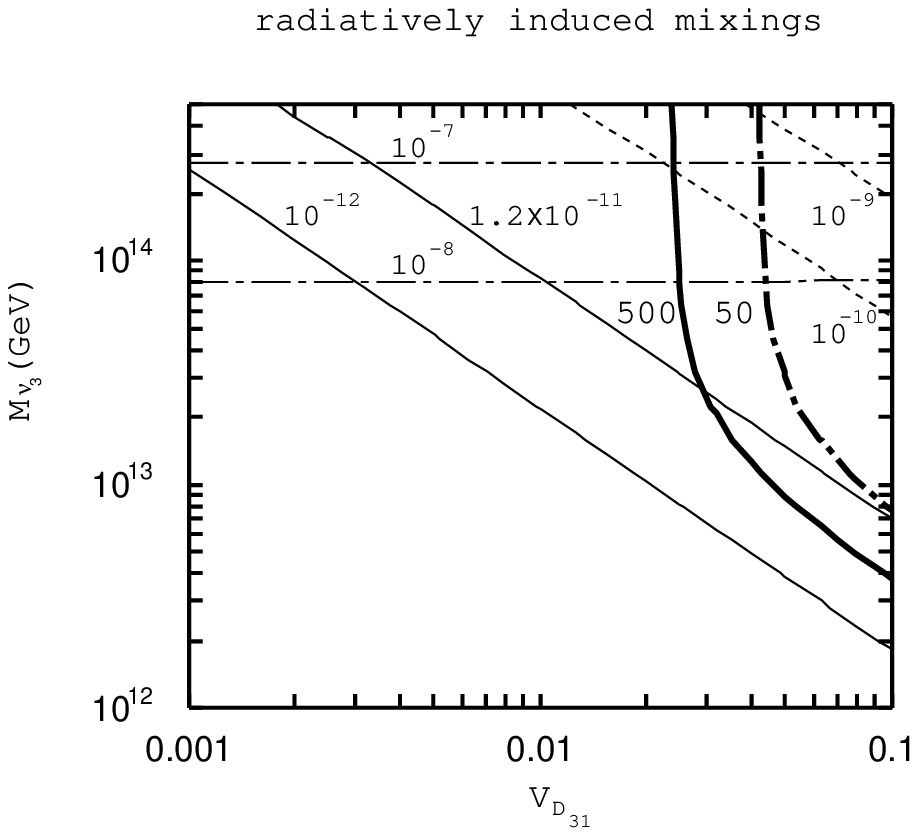,height=10cm}
\vspace*{1cm}
\caption{
Contours of $5\sigma$ discovery reaches 
in probing radiatively induced 13-mixing among left-handed sleptons 
by LC's with integrated luminosities 
${\cal L}=50  {\rm fb}^{-1}$ (thick dashed-dot contour) and 
${\cal L}=500 {\rm fb}^{-1}$ (thick solid contour)
as functions of 
the third generation right-handed Majorana mass $M_{\nu_3}$
and 
the neutrino Dirac mass term mixing matrix element $V_{D31}$.
Here we have assumed that all the off-diagonal elements 
in the left-handed slepton mass matrix are radiatively induced 
via neutrino Yukawa couplings.
In the figure we have taken $m_{\nu_\tau} = 0.07$eV and
$V_{D32} = -0.71$ 
which are suggested by the atmospheric neutrino result.
$m_{\nu_\mu}$ and $m_{\nu_e}$ are neglected.
Contours for constant branching ratios of 
$\mu \to e \gamma$ (thin solid lines), 
$\tau \to e \gamma$ (thin dotted lines), and
$\tau \to \mu \gamma$ (thin dashed-dot lines) are also shown.
Each thin solid line means 
the contour on which the branching ratio of $\mu \to e \gamma$ 
is $10^{-12}$ and $1.2 \times 10^{-11}$, respectively.
The thin dotted lines mean
the contours on which the branching ratio of $\tau \to e \gamma$ 
are $10^{-9}$ and $10^{-10}$, respectively.
The thin dashed-dot lines are
the contours of the constant branching ratio of $\tau \to \mu \gamma$, 
$10^{-7}$ and $10^{-8}$, respectively.
We take $\tan\beta=3$ in the figure 
and other SUSY parameters are shown in TABLE I.}
\label{fig:radiativereach}
\end{center}
\end{figure}

As for the 13-mixing search at LC's
we can apply the discussion on signals and BG's 
in section~\ref{sec:signalsandBGs} also to this situation,
since all the off-diagonal elements 
of the left-handed slepton mass matrix
are much smaller than the diagonal ones.
In Figure~\ref{fig:radiativereach} we show 
contours of 5$\sigma$ reach of $V_{D31}$ and $M_{\nu_3}$  
at future LC's for integrated luminosity
${\cal L} =  50{\rm fb}^{-1}$ (dashed-dot contour) and  
${\cal L} = 500{\rm fb}^{-1}$ (solid contour)
in the case all the off-diagonal elements 
in $(m^2_{\tilde{L}})$ are radiatively induced 
as in eqs.~(\ref{21element}), (\ref{31element}), and (\ref{32element}).
In the figure we take the sample parameter set of Table I
with $\tan\beta=3$ and $a_0=0$, 
and a neutrino mixing parameter set,
$m_{\nu_\tau} = 0.07 {\rm eV}$ and $V_{D32} = -0.71$,
which is consistent with the atmospheric neutrino result.
Here we have neglected $m_{\nu_\mu}$ and $m_{\nu_e}$,
assuming the hierarchy $m_{\nu_e} \ll m_{\nu_\mu} \ll m_{\nu_\tau}$.
We have used the same decay mode as that used
in section~\ref{sec:signalsandBGs} as signals.

From the figure we can see that 
for ${\cal L} =  50{\rm fb}^{-1} (500{\rm fb}^{-1})$ it may be possible 
to reach the third generation right-handed Majorana mass 
$M_{\nu_3} \gsim 10^{13}$GeV ($5\times 10^{12}$GeV) 
and the mixing matrix element $V_{D31} \gsim 0.05(0.03)$
at the statistical significance of 5$\sigma$.
Contours of constant ${\rm Br}(\tauegamma)$, ${\rm Br}(\taumugamma)$, 
and ${\rm Br}(\muegamma)$ are also shown.
We can see that for our sample parameter set 
the branching ratio of $\muegamma$ exceeds 
the current experimental bound,
Br$(\muegamma) < 1.2 \times 10^{-11}$~\cite{MEGA},
in large parameter region of 5$\sigma$ reach of LC's.
At least for our sample parameter set and our signal mode choice
near future LC's may not be so useful as $\muegamma$ in probing 13-mixing 
of the left-handed sleptons 
if there is sizable 23-mixing in the left-handed slepton mass matrix.

Some comments are in order here.

The constraint from $\muegamma$ becomes more stringent 
for moderately larger $\tan\beta$. 
This is because ${\rm Br}(\muegamma)$ 
is proportional to $\tan^2\beta$ for 
$\tan\beta \gsim 1$~\cite{LFVneutrino}, 
while our signal cross section does not change 
very much between the $\tan\beta=3$ and $\tan\beta=10$ cases
as can be inferred from Figures~\ref{fig:sigmatanbeta03} 
and~\ref{fig:sigmatanbeta10}.  
In this case the near future LC's might less useful 
than the $\muegamma$ search, at least for our sample parameter set.

Another source that can make the constraint from $\muegamma$
stronger is 12-mixing in $f_\nu$.  In the above analysis
we neglected $f_{\nu_{21}}$, which radiatively induces 
12-mixing in the left-handed slepton mass-squared matrix.
If $f_{\nu_{2}}$ and $V_{D_{21}}$ are large enough,
sizable $\muegamma$ ratio can result~\cite{HisanoNomura}.
Also in this case the reach of the near future LC's
would be less powerful than the $\muegamma$ search.

\section{conclusions}
\label{sec:conclusions}

In summary, we have calculated reach of future $\epem$ 
colliders (LC's) in probing the flavor mixing of the left-handed slepton
between the first and third generations 
by using the slepton oscillation phenomena.
$e^+e^-$ colliders are suitable for this purpose, since
they can produce, if kinematically allowed, 
sleptons of the first generation via $t$-channel
in addition to $s$-channel. 
5$\sigma$ reach of LC's  are estimated and 
we find it possible to reach 
mixing angle $\sin 2\theta_{\tilde{\nu}} \gtrsim 0.06 (0.04)$ and 
mass difference $\Delta m_{\tilde{\nu}}  \gtrsim 0.07 (0.04)$ GeV 
for sneutrinos in the first and third generations 
for our sample parameter set. 
At least for our sample parameter the slepton oscillation search 
is so sensitive that we may probe 
the flavor mixing between the first and third generations 
several orders beyond the reach of $\tautoeg$ search in near future.

We also estimated possible reach of LC's 
in probing the left-handed slepton flavor mixing
between the first and second generations which is 
severely constrained by $\mutoeg$.
The reaches of LC's correspond to Br$(\mutoeg) \simeq 10^{-12}$ 
for our sample parameters, and 
since planned $\mutoeg$ search~\cite{PSIprop} will probe Br$(\mutoeg)$ 
down to the sensitivity of $10^{-14}$, 
left-handed slepton oscillation search is not so powerful as 
$\mutoeg$ search at least for our sample parameter set.

In the case where all the off-diagonal elements coexist
in the left-handed slepton mass matrix,
especially in the case with radiatively induced sizable 23-mixing which 
is expected from the atmospheric neutrino result 
in the minimal SUGRA scenario and finite 12- and 13-mixing,
the reach of LC's in probing 13-mixing of the left-handed sleptons
can be a competition to that of $\muegamma$.
In this case, 
at least for our sample parameter set and our signal mode choice
near future LC's may not be so useful as $\muegamma$ in probing 13-mixing 
of the left-handed sleptons
if there is sizable 23-mixing in the left-handed slepton mass matrix.

\acknowledgments

The author would like to thank T. Yanagida and T. Asaka for encouragement,
K. Fujii, T. Mori, Y. Shimizu, and Y. Sumino for useful conversation, 
and is most grateful to J. Hisano and Y. Okada for comments, 
enlightening discussion, encouragement, and careful reading 
of the manuscript.
D.N. is supported by Research Fellowships of the Japan Society 
for the Promotion of Science for Young Scientists.

% \begin{figure}
% \caption{}
% \label{}
% \end{figure}

%
%%%%%%%%%%%%%%%%%%%%%%%%%%%%%%%%%%%%%%%%%%%%%%%%%%%%%%%%%%%%
%%%  FIG. 1.  %%%%%%%%%%%%%%%%%%%%%%%%%%%%%%%%%%%%%%%%%%%%%%
%%%%%%%%%%%%%%%%%%%%%%%%%%%%%%%%%%%%%%%%%%%%%%%%%%%%%%%%%%%%
%
\newpage
%
%%%%%%%%%%%%%%%%%%%%%%%%%%%%%%%%%%%%%%%%%%%%%%%%%%%%%%%%%%%%
%%%  FIG. 2.  %%%%%%%%%%%%%%%%%%%%%%%%%%%%%%%%%%%%%%%%%%%%%%
%%%%%%%%%%%%%%%%%%%%%%%%%%%%%%%%%%%%%%%%%%%%%%%%%%%%%%%%%%%%
%
\newpage
%
%%%%%%%%%%%%%%%%%%%%%%%%%%%%%%%%%%%%%%%%%%%%%%%%%%%%%%%%%%%%
%%%  FIG. 3.  %%%%%%%%%%%%%%%%%%%%%%%%%%%%%%%%%%%%%%%%%%%%%%
%%%%%%%%%%%%%%%%%%%%%%%%%%%%%%%%%%%%%%%%%%%%%%%%%%%%%%%%%%%%
%
\newpage
\vspace*{3cm}
%
%%%%%%%%%%%%%%%%%%%%%%%%%%%%%%%%%%%%%%%%%%%%%%%%%%%%%%%%%%%%
%%%                       FIG. 4.                        %%%
%%%             \label{fig:5sigmareachme}                %%%
%%% collider reach in 12-mixing vs Br(mu e gamma) reach  %%%
%%%%%%%%%%%%%%%%%%%%%%%%%%%%%%%%%%%%%%%%%%%%%%%%%%%%%%%%%%%%
%
\newpage
\vspace*{3cm}

%%%%%%%%%%%%%%%%%%%%%%%%%%%%%%%%%%%%%%%%%%%%%%%%%%%%%%%%%%%%
%%%                       FIG. 5.                        %%%
%%%              Fig.{fig:muegammadiag}                  %%%
%%%          contributions to mu to e gamma              %%%
%%%                        when                          %%%
%%%     m^2_{\tilde{L}}_{21} and m^2_{\tilde{e}}_{ij}    %%%
%%%                are negligibly small                  %%%
%%%%%%%%%%%%%%%%%%%%%%%%%%%%%%%%%%%%%%%%%%%%%%%%%%%%%%%%%%%%

%%%%%%%%%%%%%%%%%%%%%%%%%%%%%%%%%%%%%%%%%%%%%%%%%%%%%%%%%%%%
%%%                       FIG. 6.                        %%%
%%%             \label{fig:radiativereach}               %%%
%%% collider reach in radiatively induced 23-mixing      %%%
%%%             vs Br(l \to l' gamma) reach              %%%
%%%%%%%%%%%%%%%%%%%%%%%%%%%%%%%%%%%%%%%%%%%%%%%%%%%%%%%%%%%%
%
\newpage
\vspace*{3cm}

% \begin{table}
% \caption{}
% \label{}
% \end{tabular}
% \end{table}

%%%%%%%%%%%%%%%%%%%%%%%%%%%%%%%%%%%%%%%%%%%%%%%%%%%%%%%%%%%%
%%%  TABLE I %%%%%%%%%%%%%%%%%%%%%%%%%%%%%%%%%%%%%%%%%%%%%%%
%%%%%%%%%%%%%%%%%%%%%%%%%%%%%%%%%%%%%%%%%%%%%%%%%%%%%%%%%%%%

\begin{table}
\caption{The sample SUSY parameters which we used 
in our numerical calculation.
Masses are given in unit of GeV. 
We fix the lighter chargino mass ($m_{\tilde{\chi}^-_1}$) 100 GeV, 
the electron sneutrino mass ($m_{\tilde{\nu}_e}$) 180 GeV.
Other parameters are determined by solving renormalization equations
of the MSSM numerically, under an assumption of minimal SUGRA,
such as universal scalar mass at the gravitational scale 
and the GUT relation on gaugino masses.}
\begin{tabular}{lcc}
                   & $\tan\beta=3$ & $\tan\beta=10$  \\ 
\tableline
$m_{\tilde{\chi}^0_1}$ &  56  &  58 \\
$m_{\tilde{\chi}^0_2}$ & 105  & 103 \\ 
$m_{\tilde{\chi}^-_1}$ & 100  & 100 \\
$m_{\tilde{\nu}_e}$    & 180  & 180 \\
$m_{\tilde{e}_L}$      & 194  & 197 \\
$m_{\tilde{e}_R}$      & 165  & 170 \\
$\mu$                  & 244  & 200 \\
\end{tabular}
\label{MSSMparametertable}
\end{table}

%%%%%%%%%%%%%%%%%%%%%%%%%%%%%%%%%%%%%%%%%%%%%%%%%%%%%%%%%%%%
%%%  TABLE II %%%%%%%%%%%%%%%%%%%%%%%%%%%%%%%%%%%%%%%%%%%%%%
%%%%%%%%%%%%%%%%%%%%%%%%%%%%%%%%%%%%%%%%%%%%%%%%%%%%%%%%%%%%

\begin{table}
\caption{Branching ratios of three body decay of 
the lighter chargino and the second-lightest neutralino 
in sample parameter sets in TABLE I. 
(These values are taken from the work of Hisano et al.~[11].) % ~\cite{HNST}. 
}
\label{inodecaytable}
\begin{tabular}{lcc}
                       & $\tan\beta=3$ & $\tan\beta=10$ \\ 
\tableline
Br($\charginoone\to 2jets \LSP$)           &  0.63          & 0.65 \\
Br($\charginoone\to l^- \bar{\nu}_l \LSP$) & $0.12\times3$  & $0.12\times3$ \\
\tableline
Br($\neutralinotwo\to 2jets \LSP$)         &  0.23          & 0.49 \\
Br($\neutralinotwo\to l^+l^- \LSP$)        & $0.14\times 3$ & $0.09\times 3$ \\
Br($\neutralinotwo\to\nu_l\bar{\nu}_l\LSP$) 
                                           & $0.12\times 3$ & $0.08\times 3$ \\
\end{tabular}
\end{table}

%%%%%%%%%%%%%%%%%%%%%%%%%%%%%%%%%%%%%%%%%%%%%%%%%%%%%%%%%%%%
%%%  TABLE III %%%%%%%%%%%%%%%%%%%%%%%%%%%%%%%%%%%%%%%%%%%%%
%%%%%%%%%%%%%%%%%%%%%%%%%%%%%%%%%%%%%%%%%%%%%%%%%%%%%%%%%%%%

\begin{table}
\caption{Decay branching ratios of the left-handed sleptons.
We take the mass difference $\Delta m_{\tilde{\nu}} = 1 $GeV 
and the mixing angle $\theta_{\tilde{\nu}}=0$ between 
the sneutrinos of the first and the third generations.
Other parameters are the same as in TABLE I. 
(These values are taken from the work of Hisano et al.~[11].) % ~\cite{HNST}. 
}
\label{slepdecaytable}
\begin{tabular}{lcc}
                       & $\tan\beta=3$      &  $\tan\beta=10$  \\ 
\tableline
Br($\smuL (\seL) \to \mu (e) \LSP$)                 &  0.05 & 0.08 \\
Br($\smuL (\seL) \to \mu (e) \neutralinotwo$)       &  0.39 & 0.41 \\
Br($\smuL (\seL) \to \nu_\mu (\nu_e) \charginoone$) &  0.56 & 0.51 \\
Br($\stauL \to \tau \LSP$)                          &  0.06 & 0.12 \\
Br($\stauL \to \tau \neutralinotwo$)                &  0.38 & 0.38 \\
Br($\stauL \to \nu_\tau \charginoone$)              &  0.56 & 0.50 \\
\tableline
Br($\snumu (\snue) \to \nu_\mu (\nu_e) \LSP$)           &  0.30 & 0.26 \\
Br($\snumu (\snue) \to \nu_\mu (\nu_e) \neutralinotwo$) &  0.14 & 0.15 \\
Br($\snumu (\snue) \to \mu     (e)     \charginoone$)   &  0.56 & 0.59 \\
Br($\snutau        \to \nu_\tau        \LSP$)           &  0.30 & 0.26 \\
Br($\snutau        \to \nu_\tau        \neutralinotwo$) &  0.14 & 0.15 \\
Br($\snutau        \to \tau            \charginoone$)   &  0.56 & 0.59 \\
\end{tabular}
\end{table}

\end{document}